\documentclass{iaus}
\usepackage{graphicx}
\usepackage{natbib}
\def\note #1]{{\bf #1]}}
\def\titnt #1].{\hskip -0.1pt}
\def\dd{{\rm d}}
\def\bolddelta{\delta\kern-0.45em\delta\kern-0.45em\delta}
\def\boldr{\mbox{\boldmath$r$}}
\def\bolddelr{\bolddelta \boldr}
\def\muHz{\,\mu{\rm Hz}}
\def\rbf{\rm}

\title[Asteroseismology with CoRoT and Kepler] %% give here short title %%
{Stellar hydrodynamics caught in the act:\\
Asteroseismology with CoRoT and Kepler}

\author[J. Christensen-Dalsgaard \& M. J. Thompson]   %% give here short author list %%
{J{\o}rgen Christensen-Dalsgaard$^1$ \and
Michael J. Thompson$^{2,3}$}

\affiliation{$^1$Department of Physics and Astronomy, Aarhus University,
8000 Aarhus C, Denmark \\ 
email: {\tt jcd@phys.au.dk}\\[\affilskip]
$^2$School of Mathematics \& Statistics, University of Sheffield, Sheffield, S3~7RH, UK
\\[\affilskip]
$^3$High Altitude Observatory, National Center for Atmospheric Research,
P.O. Box 3000, Boulder, CO 80307-3000, USA\\
email: {\tt mjt@ucar.edu}
}

\pubyear{2011}
\volume{271}  %% insert here IAU Symposium No.
\pagerange{119--126}
\jname{Astrophysical Dynamics: from Stars to Galaxies}
\editors{N. Brummell, A.S. Brun, M. S. Miesch \& Y. Ponty, eds}
\begin{document}

\maketitle

\begin{abstract}
Asteroseismic investigations, particularly based on data on stellar oscillations
from the CoRoT and {\it Kepler} space missions,
are providing unique possibilities for
investigating the properties of stellar interiors.
This constitutes entirely new ways to study the effects of dynamic phenomena 
on stellar structure and evolution. Important examples are the extent 
of convection zones and the associated mixing and the direct and
indirect effects of stellar rotation.
In addition, the stellar oscillations themselves show very interesting 
dynamic behaviour.
Here we discuss examples of the results obtained from such investigations,
across the Hertzsprung-Russell diagram.
\end{abstract}

\firstsection % if your document starts with a section,
\section{Introduction}

%\note [Briefly on possibilities of studying stellar (magneto)hydrodynamics 
%with asteroseismology. 
%Both through studies of stellar internal structure and dynamics (rotation, 
%mainly, for now), and for the interesting dynamics of stellar pulsations
%themselves, such as large-amplitude pulsators and the excitation of modes.]
%
Stellar interiors are obvious sites for interesting dynamical phenomena,
with strong potential effects of convection and other forms of instabilities,
rotation and its evolution, as well as magnetic fields,
forming a fascinating playing field for theoretical investigations.
However, `classical' observations of stellar surfaces provide only limited
observational tests of the resulting models.
Observations of stellar oscillations, on the other hand, are sensitive to 
many aspects of the structure and dynamics of stellar interiors.
In addition, the oscillations are themselves interesting dynamical phenomena.
The diagnostic potential of oscillations
has been evident for more than a decade in the solar case,
where helioseismic investigations have yielded very
detailed information about the structure and rotation of the solar interior 
\citep[e.g.,][]{Gough1991, Christ2002} 
and the detailed properties of the solar near-surface
region through local helioseismology \citep[for a review, see][]{Gizon2010}.
For a comprehensive review of global helio- and asteroseismology, 
see also \citet{Aerts2010}.

%\note [Opportunities now opening for such studies through extensive data from
%CoRoT and {\it Kepler}.
%More to come.]

The extension of such detailed investigations to other stars has been eagerly
sought since the beginning of helioseismology, given the expected 
general presence of the oscillations in all cool stars.
There is strong evidence that the solar modes are excited stochastically
by the near-surface convection, and hence similar modes are expected in all
stars with such vigorous convective motions \citep{Christ1983}.
However, the predicted amplitudes of a few parts per million in relative
intensity or at most tens of centimeters per second in velocity have made
their detection extremely challenging.
Ground-based observations have had some success for a limited number of stars
%(for early examples, see \\note [Kjeldsen et al., Bouchy \& Carrier,
%Frandsen et al.]), 
\citep[for early examples, see][]{Kjelds1995a, Bouchy2001, Bouchy2002,
Frands2002},
but the required efforts in terms of manpower and valuable
telescope time have been very considerable.
However, in the last few years observations of stellar oscillations, and
other types of variability, have made tremendous progress, largely as a result
of the CoRoT and {\it Kepler} space missions, with significant contributions
also from the Canadian MOST satellite \citep{Walker2003, Matthe2007}.
%\note [just give a few references here, for completeness, but otherwise do
%not mention it further].
Here we provide a brief overview of some of the results from these missions
which promise to revolutionize the study of stellar internal structure
and dynamics.

\section{CoRoT and Kepler}

%\note [Brief summary of missions (with references); both target exo-planet 
%research and asteroseismology.
%Strong observational {\it and} scientific concordance with these two
%types of study, particularly exploited in {\it Kepler} (and key point 
%for PLATO).]

Both CoRoT and {\it Kepler} carry out photometric observations,
with the dual purpose of detecting extra-solar planets (exoplanets)
using the transit technique and making asteroseismic investigations.
In fact, the observational requirements for these two types of investigation
are very similar.
A central goal of the exoplanet studies is to characterize the population
of Earth-like planets in orbit around Sun-like stars.
The corresponding transit corresponds to a reduction in stellar intensity
by around $10^{-4}$.
This level of sensitivity allows the detection of solar-like oscillations in
sufficiently extended observations.
Also, both types of investigations require very long 
and continuous observations.
Finally, asteroseismology has the potential to characterize the properties
of the central stars in planetary systems detected by the transit technique.
This has proven to be very useful in a few cases for the {\it Kepler} mission
and is central to the PLATO mission proposed to ESA.
%\note [get back to both points later].

%\note [Briefly on CoRoT design, launch, field of view, operations.]

The CoRoT satellite \citep{Baglin2006, Baglin2009, Michel2008a}
was launched on 27 December 2006 into an orbit around the Earth.
The satellite has an off-axis telescope with a diameter of 28\,cm,
and a focal plane with 4 CCD detectors, two of which
(defining the exoplanet field) are optimized for 
studying planet transits and the other two, with slightly defocussed images,
optimized for asteroseismology (the asteroseismic field). 
The field of view of each CCD is $1.3^\circ \times 1.3^\circ$.
Given the low-Earth orbit, great care was taken to minimize effects of
scattered light.
Further details on the design and operations of the mission were provided by
\citet{Auverg2009}.

CoRoT's orbit allows extended observations, up to 5 months, in two regions
with a diameter of around $20^\circ$,
near the Galactic plane and the direction of
the Galactic centre and anticentre, respectively;
these are known as the `CoRoT eyes'.
The observed fields are selected within these regions, such that the
asteroseismic detectors contain a suitable sample 
of relatively bright asteroseismic targets, up to a total of 10 targets,
with an adequate density of stars in the corresponding exo field.
It should be noted that while the observations in the latter are probably not
sufficiently sensitive to study solar-like oscillations in main-sequence
stars, they do provide very extensive data on other types of stellar
variability, including solar-like oscillations in red giants
(see Section~\ref{sec:redgiant}).
The photometric analysis is carried out on the satellite, with only 
the photometric signal being transmitted
to the ground; a typical cadence for the astero field is 32\,s.

The mission suffered a loss of two of the four CCD detectors (one each
in the exo and astero fields) on 8 March 2009%
\footnote{Coincidentally a day after the launch of the {\it Kepler}
mission!}
but is otherwise performing flawlessly.
Early results on solar-like oscillations from CoRoT data were 
presented by \citet{Michel2008b}.
The mission has now been extended until March 2013.

%\note [Briefly on {\it Kepler} design, launch, field of view, operations.]

The {\it Kepler} mission \citep{Boruck2009, Koch2010}
was launched on 7 March 2009 into an Earth-trailing heliocentric
orbit, with a period of around 53 weeks.
This provides a very stable environment for the observations, in terms
of stray light and other disturbances, although with the drawback of 
a gradually decreasing rate of data transmission with the increasing distance
from Earth.
Even so, it is hoped to keep the mission operating well beyond the current
nominal lifetime of $3\frac{1}{2}$ years.
The {\it Kepler} photometer consists of a Schmidt telescope with a corrector
diameter of 95\,cm and a $16^\circ$ diameter field of view.
The detector at the curved focal plane has 42 CCDs with a total
of 95 megapixels.
This provides a field of around 105 square degrees.
The data are downlinked in the form of small images around each of
the target stars, with the detailed photometry being carried out on the
ground.
This allows up to 170,000 targets to be observed at a 30-minute cadence,
while up to 512 stars can be observed at a 1-minute cadence;
the latter are the prime targets for asteroseismology, although for
more slowly varying stars the long-cadence data are also extremely valuable.
The spacecraft is rotated by $90^\circ$ four times per orbit to keep the solar
arrays pointed towards the Sun.
Thus the observations are naturally divided into quarters.
New target lists are uploaded for each quarter, although targets can
be observed for as little as one month each;
typically, most targets are in fact observed for several quarters,
in many cases throughout the mission.
For further details on the spacecraft and the operations,
see \citet{Koch2010}.

A detector module, corresponding to two of the 42 CCD detectors, failed in
January 2010.
Otherwise, the mission has been operating successfully, reaching very close
to the expected performance.
\citet{Boruck2010} provided an early overview of {\it Kepler}
results on exoplanets.

{\it Kepler} observes a fixed field in the region of the constellations of 
Cygnus and Lyra, 
centred $13.5^\circ$ above the Galactic plane and chosen to
secure a sufficient number of targets of the right type while avoiding
excessive problems with background confusion.
A very detailed characterization of the field was carried out before
launch, resulting in the Kepler Input Catalog \citep[KIC;][]{BrownT2011}.
To avoid problems with highly saturated trailing images, the field is
located such that the brightest stars are placed in gaps between the CCDs.
In addition, the CCDs are positioned such that a star is located
at approximately the same point on a CCD following each quarterly rotation.

{\it Kepler} asteroseismology \citep[e.g.,][]{Christ2008}
is carried out in the Kepler Asteroseismic Science Consortium (KASC),
which at the time of writing has around 450 members. 
The members are organized into 13 working groups, generally dealing with 
different classes of pulsating stars.
The KASC is responsible for proposing targets, and for the analysis and 
publication of the results.
Data for the KASC are made available through the Kepler Asteroseismic Science 
Operations Centre (KASOC) in Aarhus, Denmark, which also organizes
the sharing and discussion of manuscripts before publication.
The structure of the Kepler Asteroseismic Investigation (KAI) was presented
by \citet{Kjelds2010}.

In the early phases of the KAI a survey was made of a very large number of
stars, to characterize their oscillation properties and provide the basis
for selecting targets for more extended observations.
Initial results of this survey phase were discussed by \citet{Gillil2010}.

\section{Pulsating stars}

%\note [I propose a brief section here showing the usual picture (in the b/w 
%form), possibly highlighting the cases that we are looking at 
%in the following.
%Also give the briefest summary of general pulsation characteristics, saving
%solar-like asymptotics for later.]
%
As a background for the discussion below of specific types of pulsating stars
we provide a brief overview of the properties of stellar pulsations.
For more detail, see \citet{Aerts2010}. 
We restrict the discussion to slowly rotating stars and oscillations of
modest amplitude.
In this case the oscillations can, to leading order, be characterized
as small perturbations around a spherically symmetric equilibrium structure;
individual modes depend on co-latitude $\theta$ and longitude $\phi$ as
spherical harmonics $Y_l^m(\theta, \phi)$, where $l$ measures the total number
of nodal lines on the stellar surface
and $m$ the number of nodal lines crossing the equator, with $|m| \le l$.
Modes with $l = 0$ are typically described as {\it radial} oscillations.
For each $l, m$ the star has a set of modes distinguished by the radial order
$n$.

{}From a dynamic point of view there are two basic types of stellar modes%
\footnote{In addition, modes corresponding to {\it surface gravity waves}
can be distinguished, but at degrees so far only relevant for spatially
resolved observations of the Sun.}:
acoustic modes (or p modes) where the restoring force is pressure, 
and internal gravity waves (or g modes) where the restoring force is
buoyancy variations across spherical surfaces.
Thus g modes are only found for $l > 0$.
Being acoustic 
the properties of the p modes predominantly depend on the sound-speed
variation within the star.
In many cases the result is that the frequencies $\nu$ approximately scale with
the inverse dynamical time scale, or
\begin{equation}
\nu \propto \left({G M \over R^3} \right)^{1/2} 
\propto \langle \rho \rangle^{1/2} \; ,
\label{eq:nuscale}
\end{equation}
where $M$ and $R$ are the mass and radius of the star, $G$ is the gravitational
constant and $\langle \rho \rangle$ is the mean density.
The g-mode frequencies depend on the variation of the gravitational 
acceleration and density gradient throughout the star,
the latter in turn being very sensitive to the variation of composition.

In unevolved stars typical g-mode frequencies are lower than the frequencies
of p modes.
However, as the star evolves, the core contracts, leading to a very high
local gravitational acceleration; in addition, strong composition gradients
are built up by the nuclear burning and possibly convective mixing.
In this case the local g-mode frequency may become large, and as a result the
modes can take on a mixed character, with p-mode behaviour in the outer
parts of the star and g-mode character in the deep interior.
We discuss examples of this in Sections \ref{sec:solar-like} and
\ref{sec:redgiant}.

An important measure of the properties of a mode of oscillation is its
normalized mode inertia
\begin{equation}
E = {\int_V \rho |\bolddelr|^2 \dd V \over M |\bolddelr_{\rm s}|^2} \; ,
\label{eq:inertia}
\end{equation}
where $\bolddelr$ is the displacement vector, $\bolddelr_{\rm s}$ 
the surface displacement, and the integral is over the volume $V$ of the star.
Acoustic modes have their largest amplitude in the outer layers of the star
and hence typically have relatively low inertias, whereas g modes are generally
confined in the deep interiors of stars, with correspondingly high inertia.
%\note [but this may become useful in the discussion of mixed modes.]

\begin{figure}[b]
% \vspace*{-2.0 cm}
\begin{center}
\includegraphics[width=3.5in]{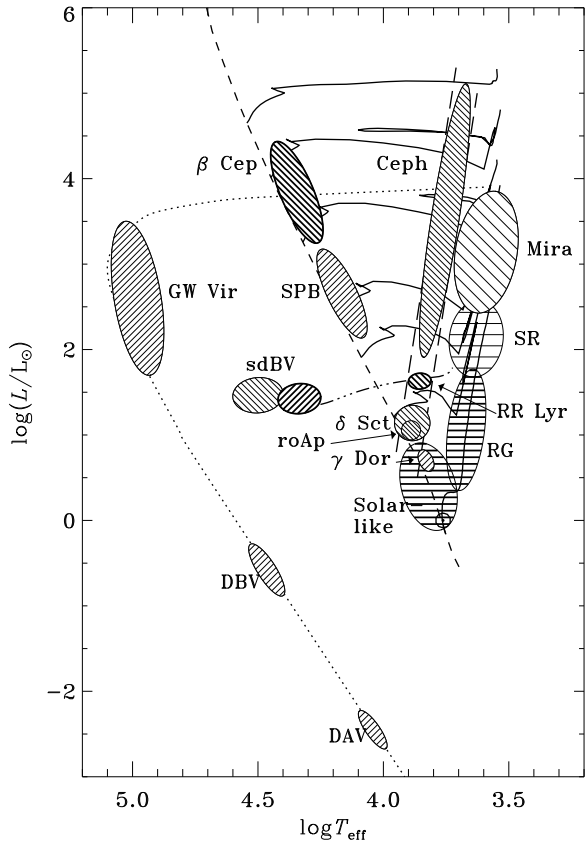} 
%\vspace*{-1.0 cm}
 \caption{Schematic location of classes of pulsating stars in the
Hertzsprung-Russell diagram.
The diagonal dashed line marks the main sequence, where stars undergo 
central hydrogen burning.
Evolution tracks following the main sequences for a few stellar masses
are shown by solid lines,
the triple-dot-dashed line indicates the location of horizontal-branch stars
with central helium burning, and the dotted line sketches the white dwarf
cooling track, after stars have stopped nuclear burning.
The hatching indicates the excitation mechanism: 
slanted lines from lower right
to upper left for heat-engine excitation of p modes, 
slanted lines from lower left
to upper right for heat-engine excitation of g modes, 
and horizontal lines for stochastic excitation.
The two nearly vertical dashed lines mark the limits of the Cepheid 
instability strip, where stars are generally excited by an opacity-driven
heat engine operating in the second helium ionization zone.
Stars discussed in the present paper are shown with bolder lines:
RR Lyrae stars (RR Lyr; Section~\ref{sec:rrlyr}),
massive main-sequence stars ($\beta$ Ceph; Section~\ref{sec:betaceph}),
long-period subdwarf B variables (sdBV; Section~\ref{sec:sdbv}),
solar-like pulsators (Section~\ref{sec:solar-like})
and red giants (RG; Section~\ref{sec:redgiant}).
%\note [Could provide more detail but that is probably not needed.]
}
   \label{fig:pulshr}
\end{center}
\end{figure}

Energetically, two fundamentally different mechanisms may excite the
oscillations.
Some stars function as a heat engine where the relative phase of compression
and heating in a critical layer of the star is such that thermal energy
is converted into mechanical energy, contributing to driving of 
the oscillations, and dominating over the other parts of the star
which have the opposite effect.
This is typically associated with opacity variations of specific
elements;
an important example is the effect of the second ionization of helium
\citep{Cox1958}, 
which causes instability in stars where the corresponding region in the star
is located at the appropriate depth beneath the stellar surface.
The driving leads to initially exponential growth of the oscillations,
with so far poorly understood mechanisms setting in to control the limiting
amplitudes of the modes.

In stars where the oscillations are not self-excited in this manner the modes
may be excited stochastically, through driving from other dynamical phenomena
in the star.
This is likely the case in the Sun where near-surface convection at nearly
sonic speed is a strong source of acoustic noise which excites the resonant
modes of the star \citep[e.g.,][]{Goldre1977, Houdek1999}.
In this case the oscillation amplitudes result from a balance between
the stochastic energy input and the damping of the modes.
Such excitation is expected in all stars with a
significant outer convection zone,
i.e., stars with effective temperature below around 7000\,K.
In principle, it excites all modes in a star; 
however, typically only modes with low inertia 
(cf.\ Eq. \ref{eq:inertia}), i.e., acoustic modes of rather high radial order,
are excited to sufficiently high amplitudes to be readily observable.

The stochastic excitation of acoustic oscillations leads to a characteristic
bell-shaped distribution of mode amplitudes
(see Fig.~\ref{fig:gemmaspec} below).
It has been found 
\citep{Brown1991, Brown1994, Kjelds1995b, Beddin2003a, Stello2008} 
that the frequency $\nu_{\rm max}$ at
maximum power scales as the acoustic cut-off frequency
\citep{Lamb1909}, leading to
\begin{equation}
\nu_{\rm max} \propto M R^{-2} T_{\rm eff}^{-1/2} \; ,
\label{eq:numax}
\end{equation}
where $T_{\rm eff}$ is the effective temperature.
This relation so far lacks a solid theoretical underpinning
\citep[see, however][]{Belkac2011},
but it has proven to be very useful in characterizing stars observed to
have solar-like oscillations (see Section~\ref{sec:redgiant}).

As illustrated in Fig.~\ref{fig:pulshr} pulsating stars are found throughout
the Hertzsprung-Russell diagram, in all phases of stellar evolution.
Thus there are excellent possibilities for learning about a broad range 
of stars.
Most of these classes have been observed by CoRoT and {\it Kepler}.
In the following we discuss a few important examples.

\begin{figure}[b]
% \vspace*{-2.0 cm}
\begin{center}
\includegraphics[width=13.5cm]{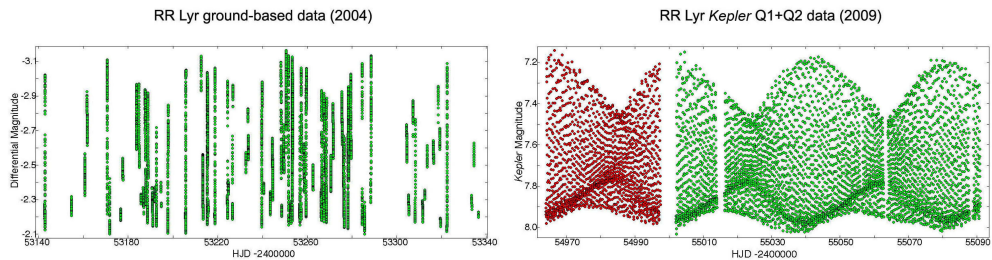} 
%\includegraphics[bb=50 50 820 350, width=5in]{RRLyr_old_new2.eps} 
%\includegraphics[width=5in]{RRLyr_old_new.eps} 
%\vspace*{-1.0 cm}
 \caption{Lightcurves for RR Lyrae lightcurves.
The left panel shows the combined results of observations
with 6 ground-based telescopes \citep{Kolenb2006}, compared with the
first two quarters of {\it Kepler} data \citep{Kolenb2011}.
%\note [There may still be problems with the figure!]
}
   \label{fig:rrlyr}
\end{center}
\end{figure}

\section{RR Lyrae stars}

\label{sec:rrlyr}
%\note [Briefly on their properties. He burning horizontal-branch stars,
%large-amplitude pulsators, excited by He II opacity heat engine mechanism.]
%
The RR Lyrae stars are amongst the `classical' pulsating stars that have been
studied extensively from the ground, since the discovery in 1900 of 
the variability of RR Lyr, the prototype of the class which 
happens to be in the {\it Kepler} field
\citep[see][for more information on these stars]{Smith1995}.
They are low-mass stars with a low content of heavy elements, in the core
helium-burning phase of evolution.
As they have relatively well-defined luminosities they serve a very useful
purpose as distance indicators to nearby galaxies, such as the Magellanic
Clouds.
Their pulsations are excited by the heat-engine mechanism,
operating as a result of
opacity variations in the second helium ionization zone.
They oscillate predominantly in one or two low-order radial modes,
with amplitudes large enough to allow observations with even very modest
equipment.

%\note [Important as distance indicators to Magellanic clouds.]

%\note [Very interesting dynamical properties, in Blazhko effect;
%also dramatic illustration of power of Kepler observations.]

What makes these stars interesting in the context of space asteroseismology
are the very interesting, and poorly understood, dynamical properties of
the oscillations in a substantial fraction of the class.
\citet{Blazko1907}%
\footnote{In the discovery paper the original Cyrillic name was
written as `Bla{\v z}ko'.
However, traditionally the name of the effect is now written as 
`the Blazhko effect' with a slightly different transliteration;
we follow that tradition here.}
discovered in one member of the class
that the maximum amplitude varied cyclically with a period of 40.8\,d.
This phenomenon has since been found in a number of RR Lyrae stars,
including RR Lyr itself \citep{Kolenb2006}.
A centennial review, including a discussion of the so far questionable
attempts at explaining the effect, was provided by \citet{Kolenb2008}.

\begin{figure}[b]
% \vspace*{-2.0 cm}
\begin{center}
\includegraphics[width=2in, angle=-90]{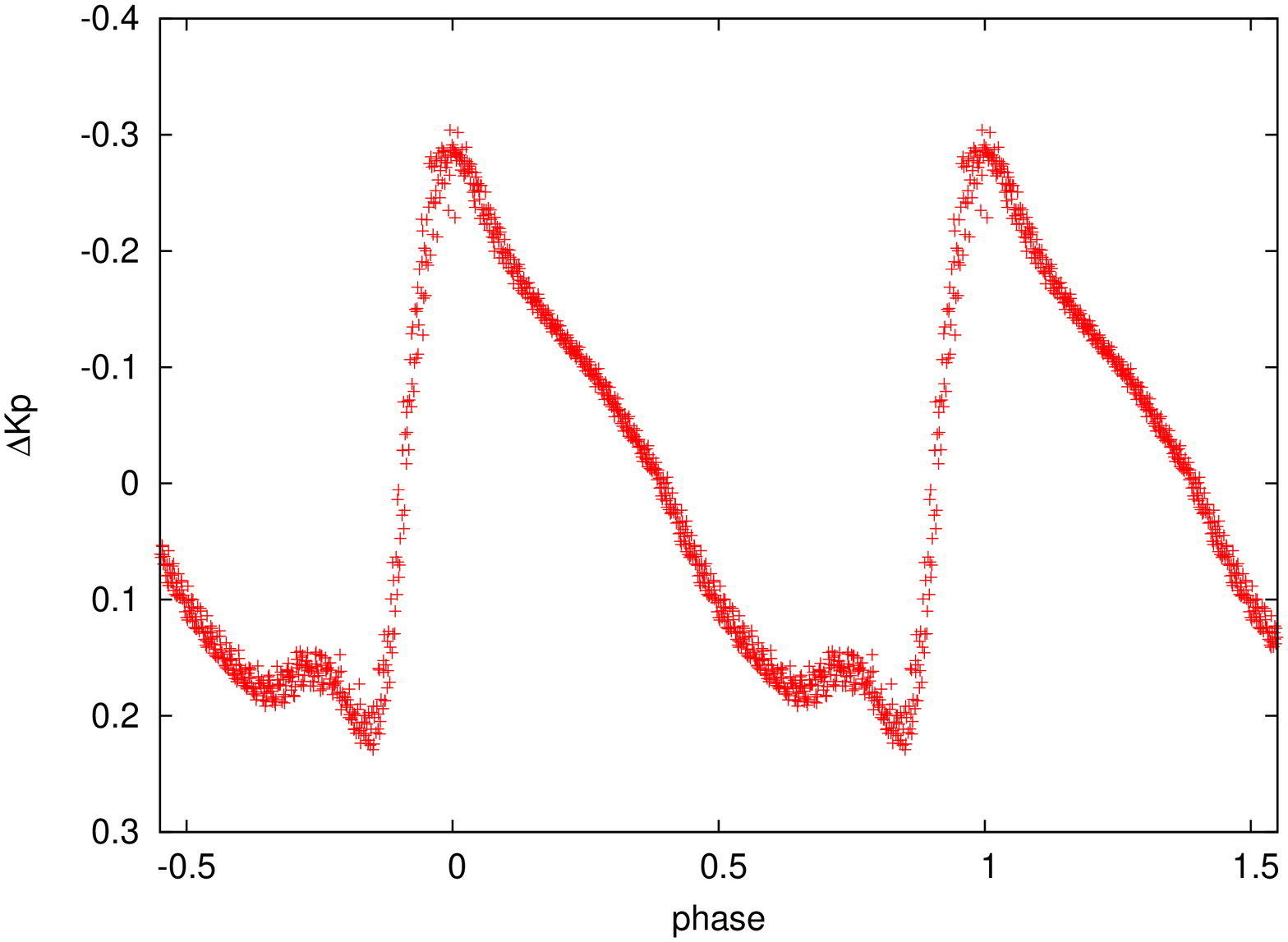} 
\includegraphics[width=2in, angle=-90]{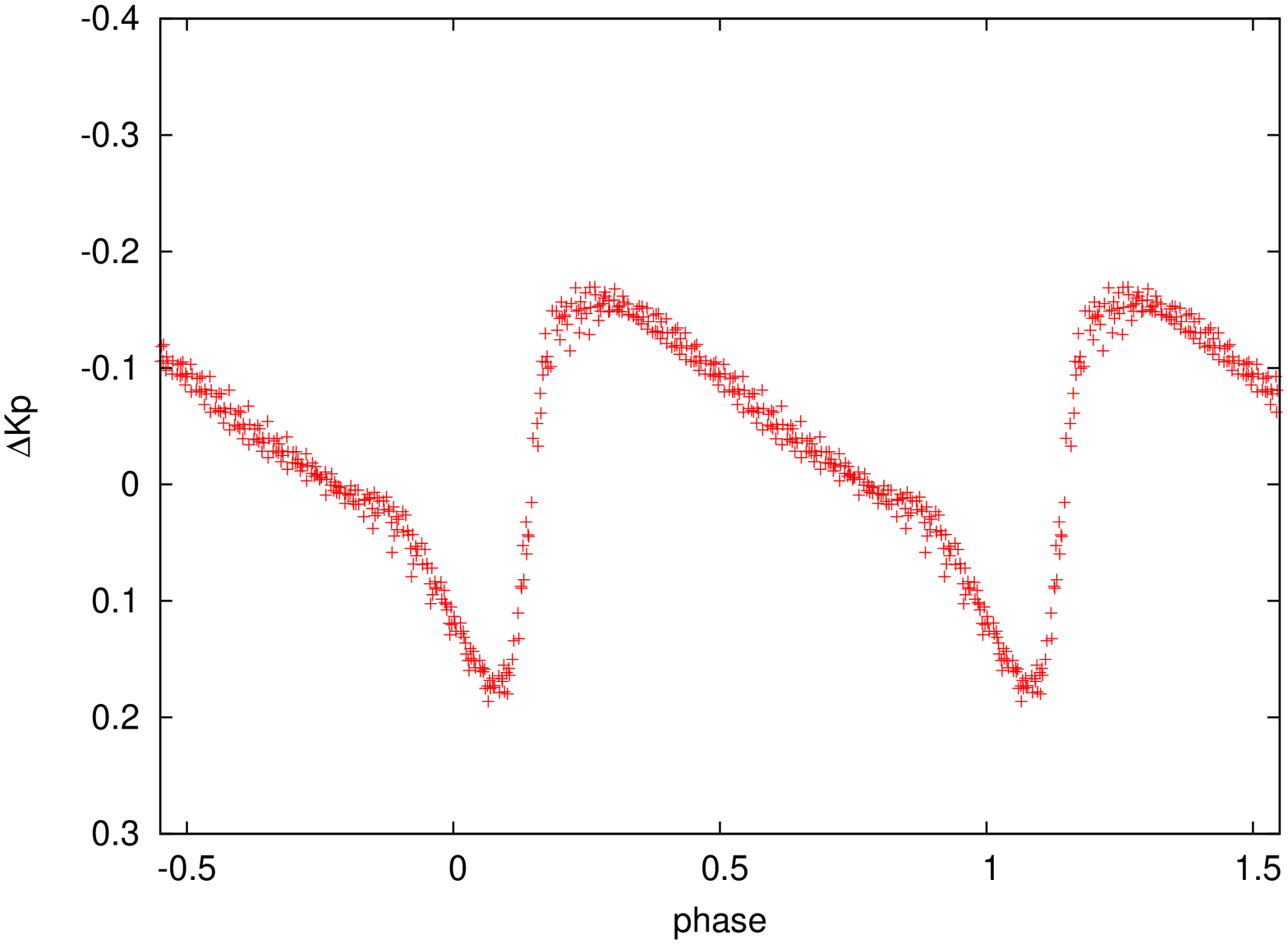} 
%\vspace*{-1.0 cm}
 \caption{Phase plots of RR Lyr at maximum and minimum amplitude,
from {\it Kepler} observations. Figure courtesy of R. Szab\'o.
%\note [Could combine into one panel.]
}
   \label{fig:rrphase}
\end{center}
\end{figure}

The continuity, duration and precision of space-based observations offer obvious
advantages in investigating such long-term phenomena.
This is illustrated in Fig.~\ref{fig:rrlyr} which compares results
of an extensive ground-based campaign on RR Lyr with {\it Kepler} observations.
The latter obviously provide a far better phase coverage throughout
the Blazhko cycle.
Phase plots at maximum and minimum amplitude in the cycle are
illustrated in Fig.~\ref{fig:rrphase}.
Similar results have been obtained by CoRoT \citep{Porett2010}.
An early survey of 28 RR Lyrae stars by {\it Kepler} \citep{Kolenb2010}
found the Blazhko effect in 40 \% of the stars, a rather higher fraction than
previously suspected.

One may hope that these vastly improved data will bring us closer
to an understanding of this enigmatic phenomenon.
As an interesting piece of evidence \citet{Szabo2010},
using {\it Kepler} data, found that 
three Blazhko RR Lyrae stars, including RR Lyr itself, showed
period doubling in certain phases of the Blazhko cycle, with slight
variations in the maximum amplitude between alternating pulsation cycles.
Also, from CoRoT observations \citet{Chadid2011} investigated cycle-to-cycle
changes in the Blazhko modulation.
Such results evidently provide constraints on, and inspiration for,
the attempts at theoretical modelling of these stars.

\section{Massive main-sequence stars}

\label{sec:betaceph}
%\note [Ideally summarize Miglio analysis, with appropriate figures?
%Important diagnostics of convective cores in supernova precursors.]
%
The pulsations in hot main-sequence stars, the so-called $\beta$ Cephei stars,
have been known for a century,
but the cause of the pulsations was only definitely identified around 1990,
when substantial revisions in opacity calculations produced opacities which
allowed excitation of the observed modes through the heat-engine mechanism
operating through the opacity from iron-group elements \citep{Moskal1992}.
This causes excitation of both p and g modes, with a tendency towards g modes
at lower effective temperature and hence mass, and a transition to the
slowly pulsating B stars (SPB stars) dominated by high-order g modes with 
very long periods.
An excellent overview of the excitation of oscillations in the B stars
was provided by \citet{Pamyat1999}.

These massive stars are of very considerable astrophysical interest as
precursors for the core-collapse supernovae.
Their structure and evolution depend strongly on the properties of the
convective core, including additional mixing outside the convectively unstable
region caused by overshoot or `semiconvection',
as well as other dynamical phenomena associated, for example, with rotation.
Thus the potential for asteroseismic investigation is very valuable, 
provided that adequate data can be obtained.
Particularly useful diagnostics can be obtained from observations of g modes.
High-order g modes are approximately uniformly spaced in period, with 
a period spacing $\Delta \Pi$ given, to leading order, by
\begin{equation}
\Delta \Pi = {2 \pi^2 \over \sqrt{l(l+1)}}
\left( \int_{r_1}^{r_2} N {\dd r \over r} \right)^{-1} \; 
\label{eq:gper}
\end{equation}
\citep{Tassou1980};
here $N$ is the buoyancy frequency and $[r_1, r_2]$ is the interval
where the modes are trapped, with $N^2 > 0$.
Assuming an ideal gas the buoyancy frequency is determined by
\begin{equation}
N^2 \simeq {g^2 \rho \over p} ( \nabla_{\rm ad} - \nabla + \nabla_\mu) \; ,
\label{eq:buoy}
\end{equation}
%$$
where $g$ is the local gravitational acceleration, $\rho$ is density and $p$
is pressure;
also, following the usual convention,
\begin{equation}
\nabla = {\dd \ln T \over \dd \ln p} \; , \qquad
\nabla_{\rm ad} = 
\left({\partial \ln T \over \partial \ln p} \right)_{\rm ad} \; , \qquad
\nabla_\mu = {\dd \ln \mu \over \dd \ln p} \; ,
\end{equation}
where $T$ is temperature, $\mu$ is the mean molecular weight and
the derivative in $\nabla_{\rm ad}$ is taken corresponding to an adiabatic
change.
In a detailed analysis \citet{Miglio2008} pointed out the diagnostic 
potential of {\it departures} from the uniform period spacing.
%in the study of the properties of convective cores.
Owing to the presence of the term in $\nabla_\mu$ in Eq.~(\ref{eq:buoy})
the buoyancy frequency is very sensitive to the detailed composition profile,
such as may result outside a convective core.
The resulting sharp features in the buoyancy frequency
%\note [should we talk about buoyancy glitches?]
introduce perturbations to $\Delta \Pi$ with a characteristic
oscillatory behaviour, the details of which depend strongly on conditions
at the edge of the core.

\begin{figure}[b]
% \vspace*{-2.0 cm}
\begin{center}
\includegraphics[width=3in]{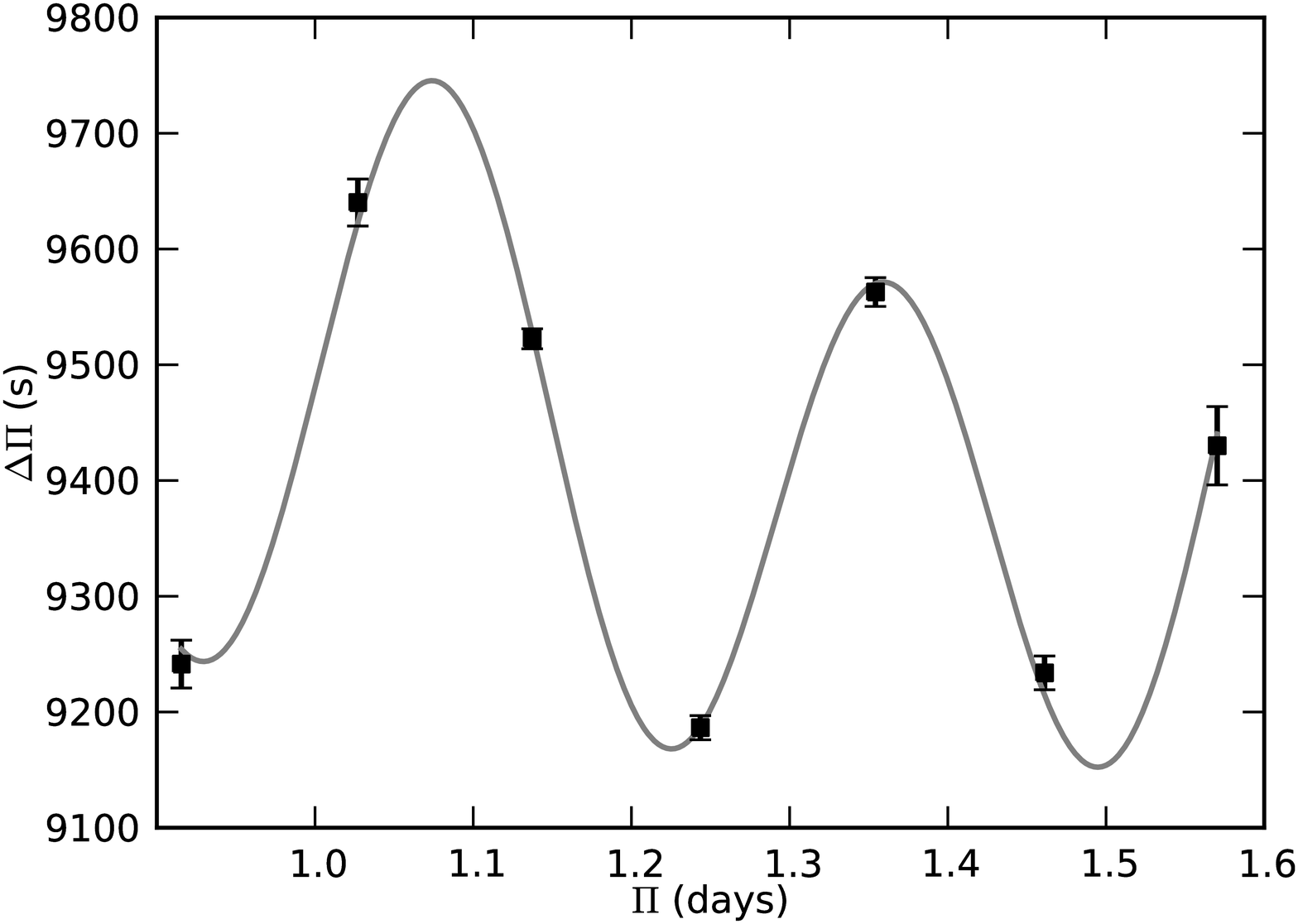} 
%\vspace*{-1.0 cm}
 \caption{Period spacings in the B3V star HD\,50230, from CoRoT observations.
The variations in $\Delta \Pi$, here fitted with a decaying sinusoid,
reflect the properties of the buoyancy frequency just outside the convective
core in the star.
Adapted from \citet{Degroo2010b}.}
   \label{fig:betaceph}
\end{center}
\end{figure}

As for the RR Lyrae stars (Section~\ref{sec:rrlyr}) the oscillations
can readily be detected in ground-based observations.
The difficulty is to obtain adequate frequency resolution and precision,
given the long periods and generally fairly dense spectra.
Substantial successes have been achieved with coordinated multi-site
observations over several months \citep{Handle2004},
leading to interesting information about convective core overshoot
\citep{Aussel2004, Pamyat2004} and internal rotation
\citep{Aerts2003, Dziemb2008}.
However, it is evident that such massive campaigns can only be carried out
in very special cases, and even then they do not provide the full desired
data continuity or sensitivity to low-amplitude modes.

%\note [Discuss Degroote et al. results, from Nature paper?]

Observations from CoRoT and {\it Kepler} have the potential to secure very 
long continuous observations of these stars \citep{Degroo2010a, Balona2011}.
A very interesting case was discussed by
\citet{Degroo2010b}, for the star HD\,50230.
This is a massive main-sequence star, of spectral type B3V, which was observed
by CoRoT for 137 days.
The resulting power spectrum showed a large number of g-mode frequencies,
with periods up to a few days, in addition to several high-frequency p modes.
In the long-period part of the spectrum the authors were able to identify a
group of eight modes with almost constant period spacing, arguing that such
a sequence is very unlikely to be found by chance.
As illustrated in Fig.~\ref{fig:betaceph} these period spacings showed a 
highly regular variation with period, of precisely the form 
predicted by \citet{Miglio2008} to result from a sharp feature in 
the buoyancy frequency.
As pointed out by Miglio et al.\ the decrease in the amplitude with increasing
period is a sensitive diagnostic of the properties of the feature.
A more detailed interpretation of the results will require more stringent
characterization of other properties of the star, through `classical'
observations as well as more extensive asteroseismic analyses of the rich
spectrum.
However, the result clearly demonstrates the potential of such observations
for characterizing the properties of convective cores in massive main-sequence
stars.
 
%\begin{table}
%  \begin{center}
%  \caption{Overview of current knowledge on circum-stellar condensate grains in meteorites.}
%  \label{tab1}
% {\scriptsize
%  \begin{tabular}{|l|c|c|c|c|}\hline 
%{\bf Mineral} & {\bf Size [$\mu$m]} & {\bf Isotopic Signatures} & {\bf Stellar} & {\bf Contri-} \\ 
%   &  {\bf abund.}  [ppm]$^1$ & & {\bf Sources} & {\bf bution$^2$} \\ \hline
%diamond & $~0.0026$ & Kr-H, Xe-HL, Te-H & supernovae & ? \\
% nitride & $~ 0.002$ & & & \\ \hline
%  \end{tabular}
%  }
% \end{center}
%\vspace{1mm}
% \scriptsize{
% {\it Notes:}\\
%  $^1$For the abund.\ (in wt.\ ppm) the reported maximum values from different meteorites are given. \\
%  $^2$Note uncertainty about actual fraction of diamonds that are pre-solar and for fraction of graphite attributed to SN and AGB stars (see discussion in text).}
%\end{table}

\section{Subdwarf B stars}

\label{sec:sdbv}
%\note [Briefly on character, likely evolutionary route (importance of binary
%evolution)?]
%
The subdwarf B stars (sdB stars) are very hot core helium burning stars,
at the blue end of the horizontal branch \citep[for a review, see][]{Heber2009}.
The high effective temperature is the result of the stars having lost most 
of their hydrogen envelope, through processes that are so far not fully
understood.
Pulsations were first observed in such stars by \citet{Kilken1997},
with very high frequency.
That the stars might be unstable to acoustic modes was found in parallel,
and independently, by \citet{Charpi1996}.
The driving arises from the heat-engine mechanism operating on opacity 
from the iron-group elements.
Subsequently \citet{Green2003} also observed long-period oscillations in
somewhat cooler sdB stars, corresponding to g modes of high order.
A detailed analysis of the excitation was carried out by \citet{Fontai2003}.
To be effective, the iron-group opacity must be enhanced through
enrichment of the elements in the critical region through radiative 
levitation \citep[see also][]{Fontai2006}; 
owing to the high gravitational acceleration such processes of settling
and levitation are quite efficient in these stars.

\begin{figure}[b]
% \vspace*{-2.0 cm}
\begin{center}
\includegraphics[width=2in]{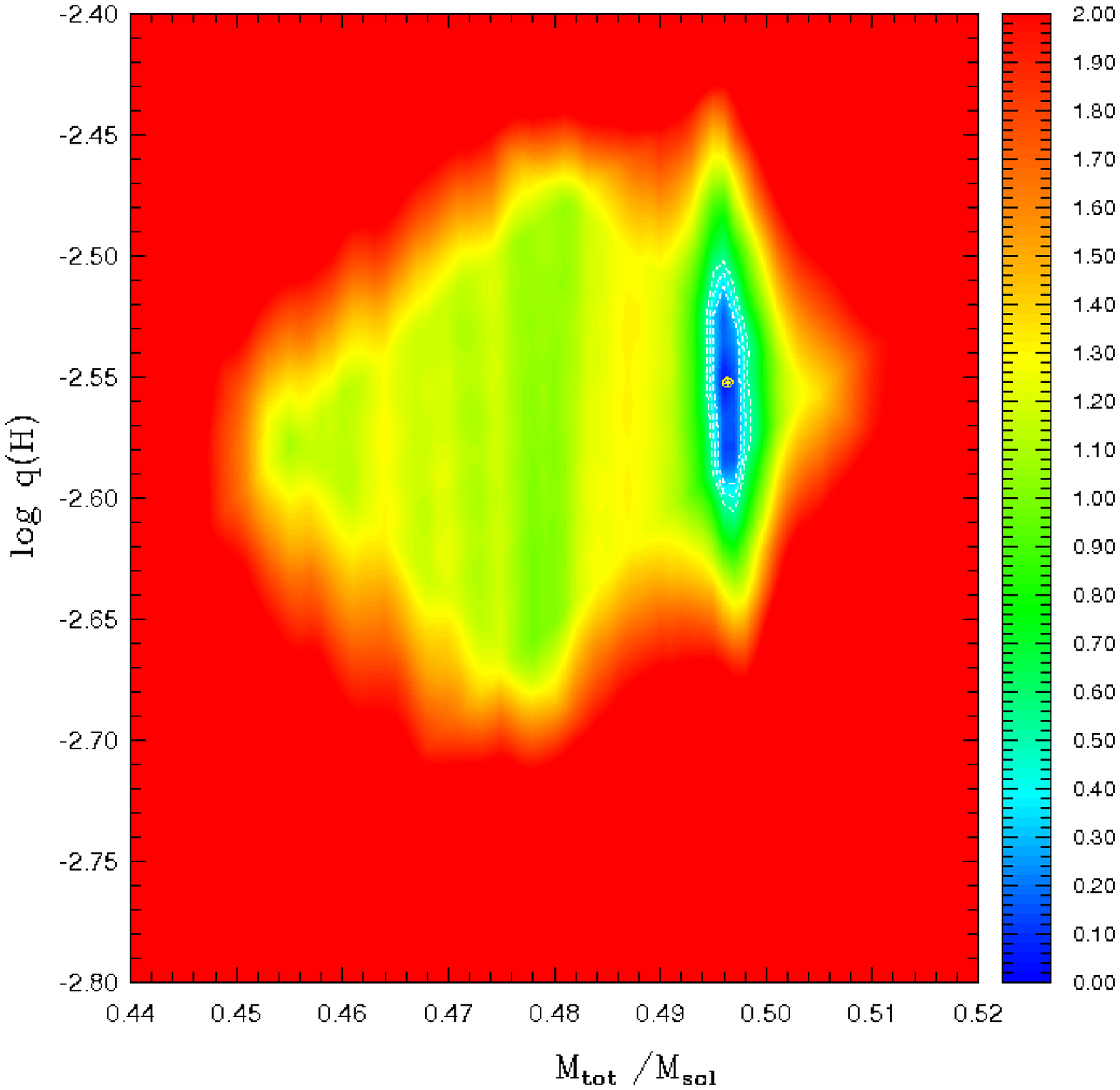} 
\includegraphics[width=2in]{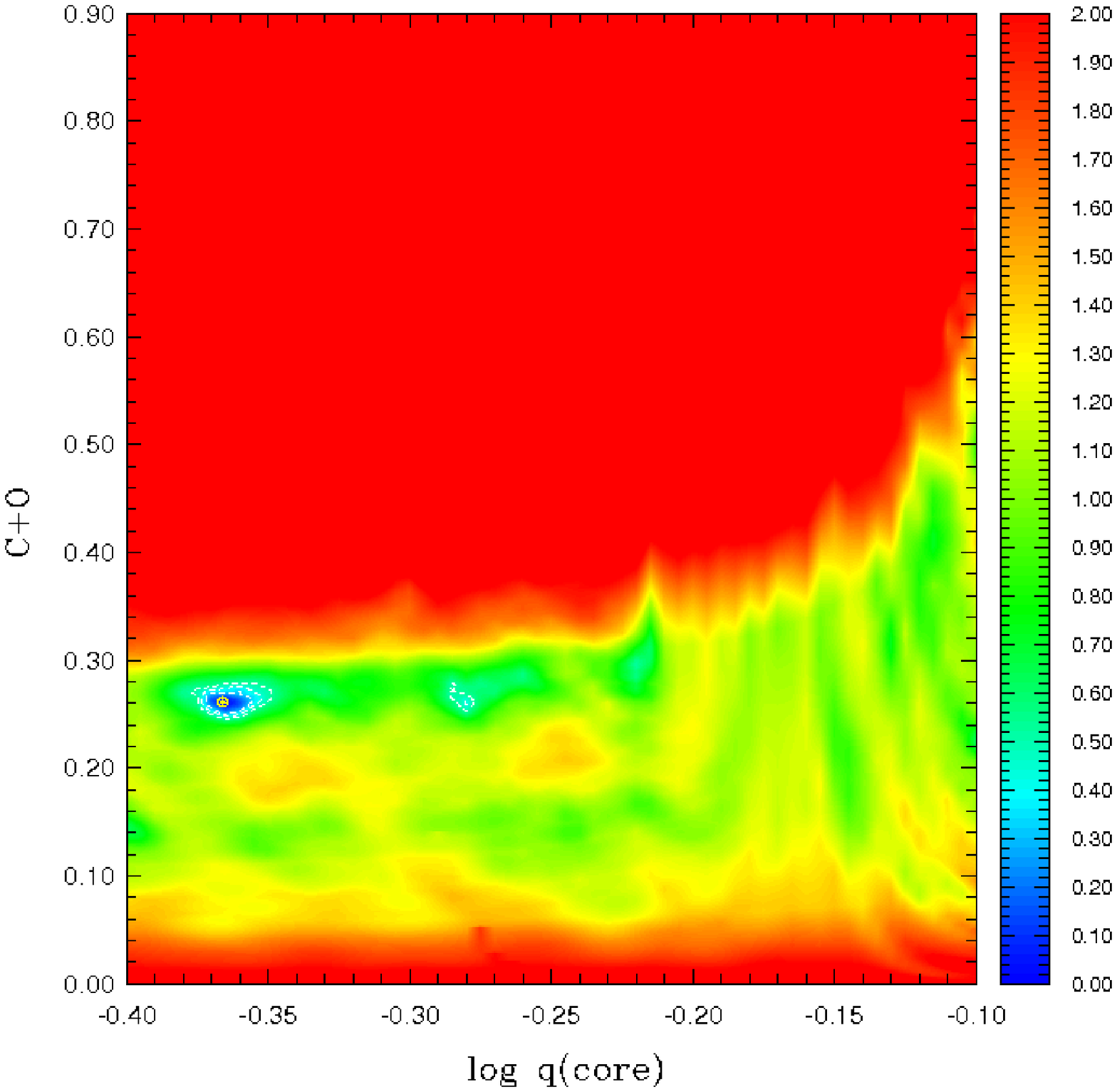} 
%\vspace*{-1.0 cm}
 \caption{The merit function $S^2$ (cf. Eq.~\ref{eq:sdbfit}) in a fit 
of the observed {\it Kepler} periods for the star KPD~1943+4058 to a set 
of models; the colour scale for $\log S^2$ is indicated to the right.
The left panel shows the fit in terms of the total mass of the star, in solar
units, and the logarithm of the fraction of the mass in the hydrogen-rich
envelope. 
The right panel plots the merit as a function of the logarithm of the mass 
fraction outside the mixed core and the combined abundance, by mass, of
carbon and oxygen in the core.
{}From \citet{VanGro2010a}.}
   \label{fig:sdbchi}
\end{center}
\end{figure}

%\note [Very briefly on discovery history, including now also `Betsy stars',
%(need to check proper nomenclature!). Emphasize difficulty of observing 
%long-period multimode oscillators from the ground.]

As in the previous cases discussed, major ground-based efforts have been 
made to study these pulsations, involving coordinated observations between
several observatories over extended periods
\citep[e.g.,][]{Randal2006a, Randal2006b, Baran2009}.
The difficulties of such observations are particularly severe for the 
g-mode pulsators, with periods of order hours and amplitudes of only a 
few parts per thousand.
Yet these modes are particularly interesting as diagnostics of the 
deep interiors of the stars.
Thus the sdB variables have been prime targets for asteroseismology
with CoRoT and {\it Kepler}.
The {\it Kepler} initial survey has included a substantial number of sdB stars
and so far led to the discovery of 14 pulsating stars,
all except one of which are long-period pulsators, with substantial numbers
of g modes of intermediate and high order
\citep{Ostens2010, Ostens2011, Baran2011}.
Thus the periods may be expected to satisfy the asymptotic relation
(\ref{eq:gper}).
\citet{Reed2011} showed that this is indeed the case and noted the importance
for mode identification.

As a specific example of the power of these data
we consider a detailed analysis of the star 
KPD~1943+4058 based on the initial {\it Kepler} data.
Here \citet{Reed2010} detected 21 modes, with periods between 0.7 and 2.5\,h,
and in addition three modes in the p-mode region.
Similar observational results were obtained by \citet{VanGro2010a},
who in addition carried out a fit of the g-mode periods to models of the star.
The models were described by their total mass, the mass in the thin 
envelope and an assumed fully mixed core, and the composition of the core,
characterized by the combined abundance of carbon and oxygen.
In addition, the models were constrained to be consistent with the
spectroscopically inferred effective temperature and surface gravity.
The fit to the observations measured by a merit function $S$ defined by
\begin{equation}
S^2 = \sum_i (\Pi_i^{\rm (obs)} - \Pi_i^{\rm (mod)} )^2 \; ,
\label{eq:sdbfit}
\end{equation}
where $\Pi_i^{\rm (obs)}$ and $\Pi_i^{\rm (mod)}$ are the observed and
model periods, respectively, and the identification of the computed modes
with the observed modes is part of the fitting procedure.
Some results are illustrated in Fig.~\ref{fig:sdbchi}.
The analysis provided very precise determinations of the properties of the
star, including strong evidence for a mixed core region significantly larger
than the expected convectively unstable region, indicating substantial core
overshoot.

It should be noted that the periods of the best-fitting 
model did not agree with
the observed periods to within their observational uncertainty.
This evidently shows that further improvements of the modelling, beyond
the parameters included in the fit, are needed.
It seems likely that the number of observed periods is sufficient that a 
formal inversion of the differences can be attempted, as has been 
applied with great success in the solar case \citep[e.g.,][]{Gough1996}.
The results may provide a direct indication of the aspects of the models 
where improvements are needed.

Similar analyses will be possible for the remaining stars for which 
extensive g-mode data have been obtained with {\it Kepler}, and they will
evidently improve as the stars continue to be observed.
Also, an sdB star showing extensive g-mode pulsations has been observed
by the CoRoT mission \citep{Charpi2010}.
Model fitting to the resulting periods by \citet{VanGro2010b} yielded
results rather similar to those discussed above.

%\note [Probably just show $\chi^2$ plot, emphasize detection of evidence for
%convective-core overshoot.]

\section{Solar-like oscillations in main-sequence stars}

\label{sec:solar-like}
%\note [Stochastically excited by convection.
%Very briefly on properties: asymptotics, $\nu_{\rm max}$, lifetime and
%amplitude.]
%\note[Stochastic excitation; Tassoul formula; Echelle diagram; CD diagram.
%Something about $\nu_{\rm max}$, lifetime and amplitude.] 
Solar-like oscillations are predominantly acoustic in nature and excited by 
turbulent convection in the star's outer convective envelope. As already 
noted, although this broadband excitation mechanism excites all modes in 
principle, because of their low mode inertias it tends to be the high-order
p modes that are excited to observable amplitude. The first star in which such 
oscillations were detected was of course the Sun, and the study of the Sun's 
oscillations has led to the rich field of helioseismology, in which Juri Toomre
has played a leading role \citep[see, {\it e.g.},][]{Gough1991, Christ2002}.

\begin{figure}[b]
% \vspace*{-2.0 cm}
\begin{center}
\includegraphics[width=4.5in]{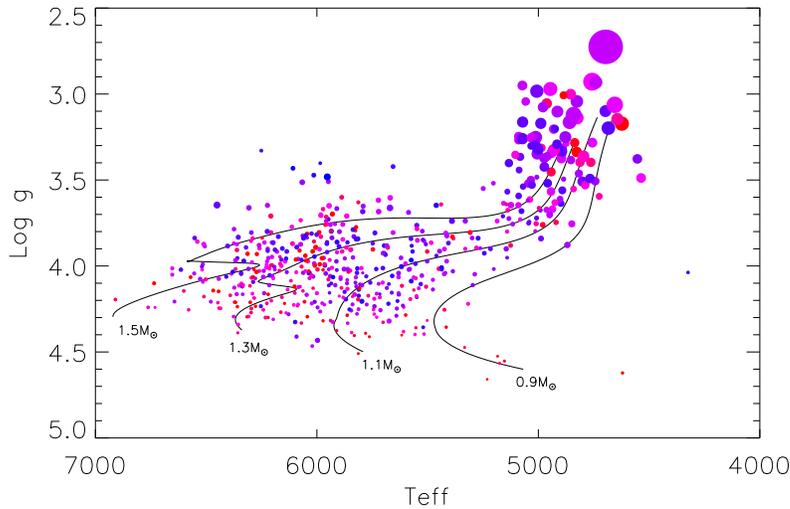} 
%\vspace*{-1.0 cm}
 \caption{
Location of stars observed by {\it Kepler} to show solar-like oscillations,
plotted as a function of effective temperature and the logarithm of the surface
gravity.
The size of the symbols measures the oscillation amplitude, while the colour
(in the electronic version)
indicates the apparent brightness of the stars, with red for the
brightest stars.
For comparison, evolution tracks at the indicated masses are also shown.
Adapted from \citet{Chapli2011} and \citet{Verner2011}.
Figure courtesy of C. Karoff.
%\note [Location of solar-like pulsators in 'HR' diagram.
%Size indicates amplitude, colour Kepler magnitude (red is brightest).
%Refer to \citet{Verner2011}. Thanks to C. Karoff for figure.]
}
   \label{fig:solar-like}
\end{center}
\end{figure}

The asteroseismic investigation of solar-type stars has taken a major step 
forward thanks to the {\it Kepler} mission \citep{Chapli2010}.
To date, {\it Kepler} has yielded
clear detections of oscillations in 500 solar-type stars \citep{Chapli2011}.
A plot of
the distribution in the HR diagram of {\it Kepler} stars with detected solar-like
oscillations is shown in Fig.~\ref{fig:solar-like}.
This represents an increase by more than a factor of 20 of the number of
known stars on and near the main sequence which show solar-like oscillations.

The high-order, low-degree p modes in solar-type stars
occupy resonant acoustic cavities that extend from the surface to
a region close to the stellar core.
Their cyclic frequencies $\nu_{nl}$ satisfy a simple expression:
\begin{equation}
\nu_{nl} \simeq \Delta \nu \left( n + {l \over 2} + \epsilon \right) 
- d_{nl} \; .
\label{eq:pasymp}
\end{equation}
Here 
\begin{equation}
\Delta \nu = \left(2 \int_0^R {\dd r \over c} \right)^{-1} \; ,
\label{eq:largesep}
\end{equation}
where $c$ is the adiabatic sound speed and the integral is over the
distance $r$ to the centre of the star;
also
\begin{equation}
d_{nl} = {l(l+1)\Delta\nu\over 4\pi^2\nu_{nl}}
\left[
{c(R)\over R} - \int_0^R{{\rm d}c\over{\rm d}r}{{\rm d}r\over r}
\right]
\end{equation}
\citep{Tassou1980, Gough1986}. 
%\note[also Gough in the Unno festscrift].
Accordingly, the frequencies of such modes of the same degree are separated by 
{\it large separations}
\begin{equation}
\Delta \nu_{nl} = \nu_{n l} - \nu_{n\, l-1}
\approx \Delta\nu\;,
\label{eq:large_sep}
\end{equation}
while the small correction $d_{nl}$
gives rise to 
{\it small separations} 
\begin{equation}
\delta \nu_{nl} = \nu_{nl} - \nu_{n-1 \, l+2}
\approx
(4l+6){\Delta\nu\over 4\pi^2\nu_{nl}}
\left[
{c(R)\over R} - \int_0^R{{\rm d}c\over{\rm d}r}{{\rm d}r\over r}
\right]
\end{equation}
between the frequencies of modes that differ in degree by 2 and in order by 1.
Finally $\epsilon$ is a slowly varying function of frequency which
is predominantly determined by the properties of the near-surface region.

\begin{figure}[b]
% \vspace*{-2.0 cm}
\begin{center}
\includegraphics[width=4in]{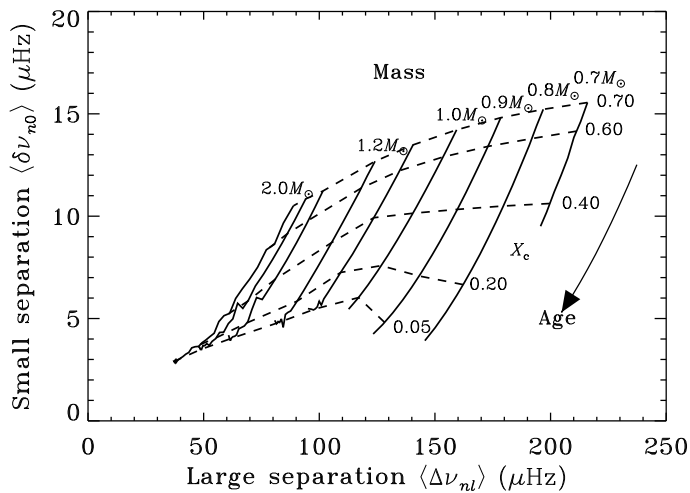}
%\vspace*{-1.0 cm}
 \caption{Asteroseismic HR diagram for a homogeneous set of 
stellar models
of different masses and ages. The solid lines show how a star of given mass
evolves in terms of its large and small frequency separations. The dashed
lines connect stars which have aged by the same fraction of their total
main-sequence lifetime.  
%\note[Need somehow to relabel these axes or otherwise
%reconcile the notation for mean large and small separations.]
%\note [Will be done; that should be straightforward.]
}
   \label{fig:asteroHR}
\end{center}
\end{figure}

The quantity $\Delta\nu$ is a measure of the acoustic radius of the star. It
shares the scaling (Eq.~\ref{eq:nuscale}) of the frequencies
with the mean density, and hence so too do the large separations.
For main-sequence stars $d_{nl}$, and thus the small separations, are
mainly determined by the central regions of the star, being 
sensitive in particular to the sound-speed gradient in the core, 
and hence they provide
a measure of the star's
evolutionary state. Thus measuring the large and small 
frequency separations gives a measure of the mean density and evolutionary 
state of the star. 
A useful seismic diagnostic is the 
asteroseismic HR diagram, in which the star's average large separation is plotted
against its average small separation: this is illustrated in 
Fig.~\ref{fig:asteroHR}.
For main-sequence stars, the asteroseismic HR diagram allows the mass and 
age of the star to be estimated, assuming that other physical inputs (such as 
initial chemical composition and the convective mixing-length parameter) are 
known.

\begin{figure}[b]
% \vspace*{-2.0 cm}
\begin{center}
\includegraphics[width=3in]{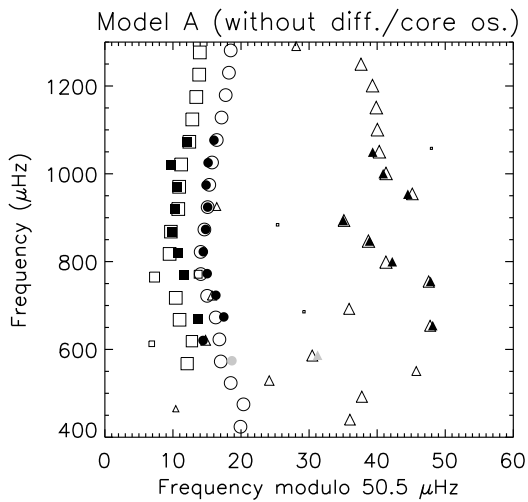} 
%\vspace*{-1.0 cm}
 \caption{An \'echelle diagram for the solar-type star KIC~11026764 
(Gemma) indicating both the observed frequencies (filled symbols) and 
those of a stellar model (open symbols). 
Modes of degree $l=0, 1, 2$ are denoted respectively by circles, triangles
and squares. From \citet{Metcal2010}.}
   \label{fig:gemmaechl}
\end{center}
\end{figure}

The existence of the large separation also motivates another diagnostic, which
is to plot the frequencies of the star in a so-called {\it \'echelle diagram}.
Here the frequencies $\nu_{nl}$ are reduced modulo $\Delta\nu$ and 
$\bar\nu_{nl} \equiv \nu_{nl}\rm{\ mod\ }\Delta\nu$ is
plotted on the x-axis while $\nu_{nl}-\bar\nu_{nl}$ is plotted on the y-axis.
If the spacing of frequencies according to asymptotic expression
(\ref{eq:pasymp}) were exact, the modes of like degree would be aligned as 
nearly vertical lines in the \'echelle diagram, with the lines corresponding to 
$l=0,2$ being separated from one another by the small separation and
the line corresponding to $l=1$ being offset from those by an
amount corresponding to half the large separation. Deviations from such a 
simple picture reveal deviations from the simple asymptotic relation and 
contain physically interesting information about the star. An example of an 
\'echelle diagram for star KIC~11026764, is shown in Fig.~\ref{fig:gemmaechl}. 
The ridges corresponding to $l=0,2$ are evident. 
The ridge corresponding to $l=1$
is more irregular: this is due to avoided crossings in this relatively 
evolved star, an issue which is discussed further below.

\begin{figure}[b]
 %\vspace*{-1.5 cm}
\begin{center}
\includegraphics[width=4.5in]{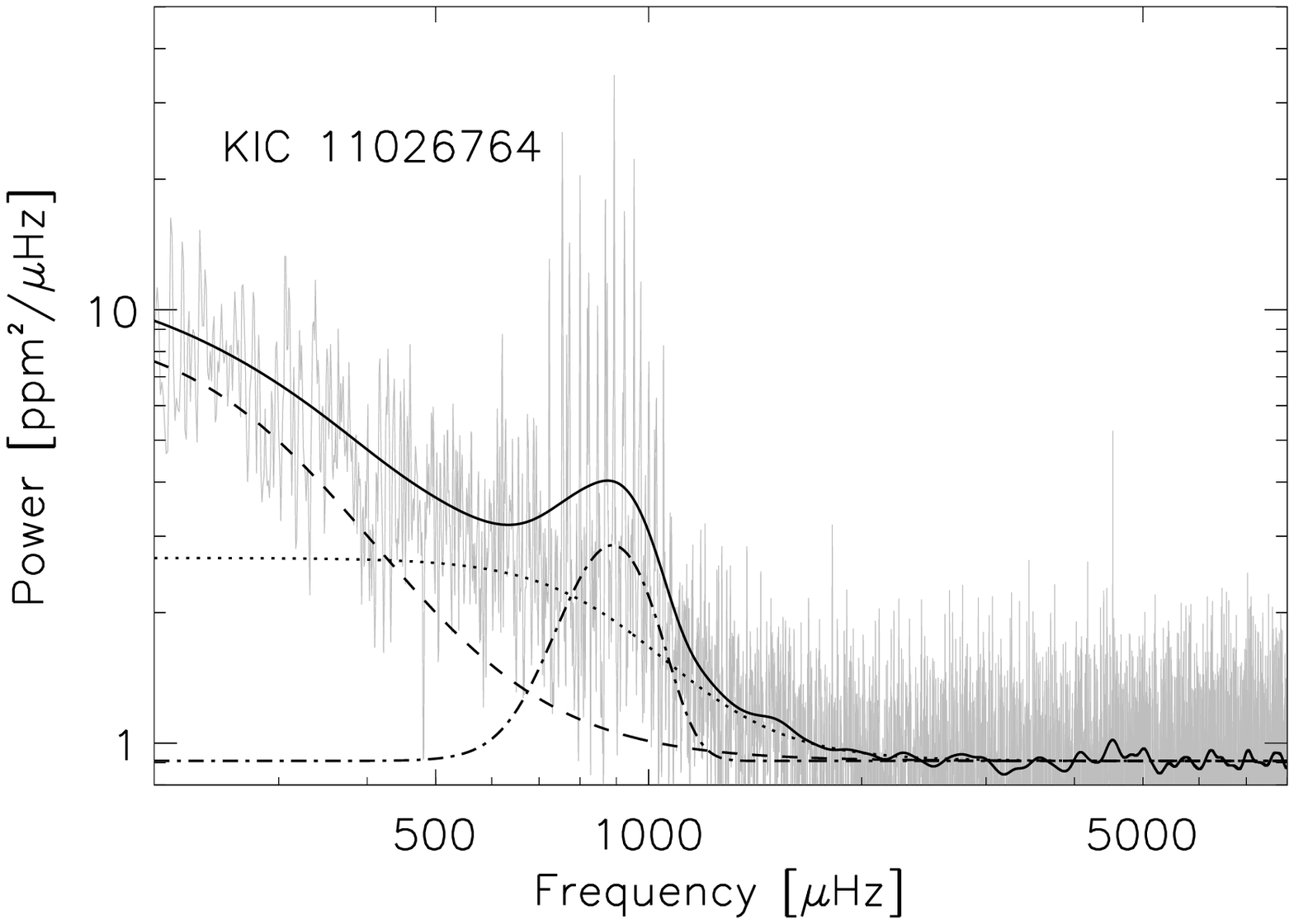} 
%\vspace*{-2.0 cm}
 \caption{Observed power spectrum of the star KIC~11026764, known as
Gemma, based on early {\it Kepler} observations.
The solid line shows the smoothed spectrum, separated into the oscillation
signal (dash-dotted line) and background components from granulation and
faculae (dashed and dotted lines, respectively).
Figure, adapted from \citet{Chapli2010}, courtesy of C. Karoff.
}
   \label{fig:gemmaspec}
\end{center}
\end{figure}

Even without a measurement of the small separations, it is possible to 
make useful seismic estimates of stellar masses and radii using observational
measures of $\Delta \nu$ and of the frequency $\nu_{\rm max}$ of maximum
mode amplitude. From Eq.~(\ref{eq:largesep}) the former scales as $M/R^3$, whereas by 
Eq.~(\ref{eq:numax}) the latter scales as $M/R^2$: hence a measurement of these
two yields an estimate of both $M$ and $R$.  This has been applied to an 
ensemble of 500 
stars in the {\it Kepler} field by \citet{Chapli2011}. 
This paper by Working Group 1 (WG1) of the KASC 
concludes that while the
estimated radii of the 500 stars
are similar to those expected from stellar population models,
the observed distribution of masses is
wider at its peak than the model distribution, and is offset towards slightly 
lower masses.

\citet{Chapli2010} published observations of three bright solar-type stars,
which were monitored during the first month of {\it Kepler}
science operations. This paper was the first
to establish the asteroseismic potential 
of {\it Kepler} observations of solar-type stars:
about 20 modes were distinguished 
in each star, and the frequencies and frequency separations allowed radii, 
masses and ages of the stars to be estimated. The three stars that were 
the objects of the study, KIC~6603624, KIC~3656476 and KIC~11026764, were 
given the working names Saxo, Java and Gemma%
\footnote{The working group allocated the names of pet cats to the stars that have been the early objects of study: this ideosyncracy is due to the WG1 lead, 
Bill Chaplin.}
by the WG1 members.  One of these stars, Gemma, was revealed to have evolved 
off the main sequence 
and proved more challenging to model and constrain asteroseismologically: 
this interesting case is discussed further now.

Gemma is one of the best-studied solar-like stars to be investigated
thus far with {\it Kepler} data.
The observed power spectrum, as obtained by \citet{Chapli2010},
is shown in Fig.~\ref{fig:gemmaspec}.
The analysis of the frequencies of the star was the subject of the paper by 
\citet{Metcal2010}. Gemma is 10-20 per cent more massive than the Sun and also
somewhat older. The core of Gemma is therefore more chemically evolved than is
the core of the Sun;
the models indicate that the star has a small compact helium core,
surrounded by a hydrogen-burning shell.
This leads to interesting behaviour of the
star's frequencies which provides a powerful diagnostic of the star's 
evolutionary state. As a solar-like star evolves at the end of its main-sequence
life, it continues to grow in radius while forming a strong gradient in 
mean-molecular weight at the edge of its core.
Also, the core contracts, increasing the central condensation and
the gravitational acceleration in the deep interior of the star.
These effects in turn cause a strong peak to form in the buoyancy frequency
(cf.\ Eq.~\ref{eq:buoy}),
which supports g modes at frequencies which are representative for the
stochastically excited solar-like oscillations.
Thus at such frequencies the star has two resonant cavities supporting 
nonradial modes:
one in the deep envelope of the star where the modes behave like p modes,
increasingly confined to the outer regions of the star with increasing degree,
and one in the core where the modes behave like g modes.
These two regions are separated by an intermediate region where the modes
are evanescent.
%\note[Should mention the central 
%condensation and increased gravitational acceleration also.]

\begin{figure}[b]
% \vspace*{-2.0 cm}
\begin{center}
\includegraphics[width=12cm]{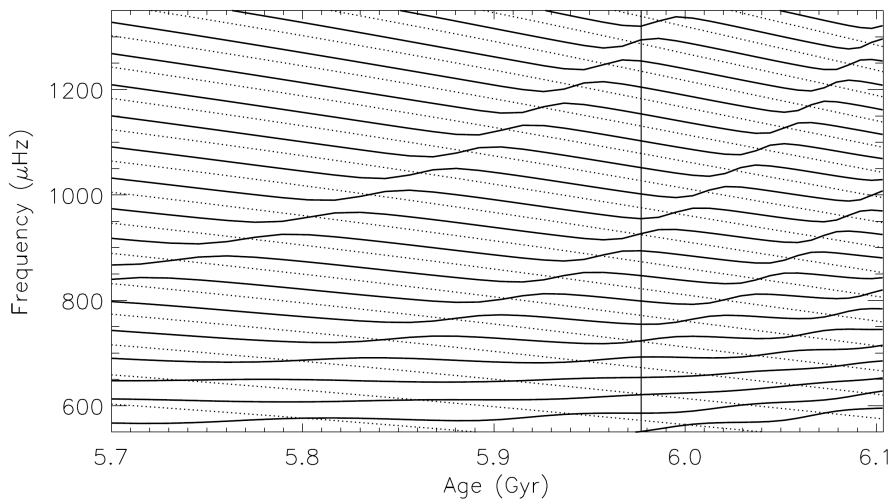} 
%\vspace*{-1.0 cm}
 \caption{Evolution of the $l=0$ (dotted) and $l=1$ (solid) mode frequencies
as a function of age for a representative stellar model of 
KIC~11026764 (Gemma). The vertical line indicates the age of 5.77\,Gyr of 
one good-fitting model to Gemma's observed frequencies.
{}From \citet{Metcal2010}.
   \label{fig:gemmaevol}}
\end{center}
\end{figure}

With increasing age the star undergoes an overall expansion which
causes the frequencies of the p modes to decrease, while the increase in
the central condensation and hence the
buoyancy frequency causes the frequencies of g modes to increase. Although the 
g modes are not in general
themselves observable directly, at times in the star's 
evolution the frequencies of a g mode and a p mode get sufficiently close 
for a strong coupling between the two modes to be possible, giving rise to a 
so-called mixed mode. This evolution of the frequency spectrum is illustrated
in Fig.~\ref{fig:gemmaevol} for a representative stellar model of Gemma. 
This shows how the radial ($l=0$) and $l=1$ p-mode frequencies change as 
a function of age of the star. By their nature, g modes are non-radial; and 
hence the radial modes cannot couple to them: thus the overall expansion of the
star simply causes the $l=0$ frequencies to decrease monotonically with 
increasing age. The $l=1$ frequencies also tend to decrease with age; but 
occasionally a given $l=1$ mode approaches the frequency of an $l=1$ g mode: 
at that point, a strong coupling between the two modes occurs if the
evanescent region between their two resonant cavities is not too large. 
The frequencies of the 
%p and g mode 
two modes
never actually cross: instead, the modes
undergo an avoided crossing, which results in the observable mode increasing
in frequency as the star evolves, for the duration of the strong mode
coupling. 
%Although the g-mode frequencies are not plotted in 
% Fig.~\ref{fig:gemmaevol}, the g-mode ridges 
The evolving frequencies of the physical g~modes can 
%nonetheless 
be discerned
in Fig.~\ref{fig:gemmaevol}
as the loci of the avoided crossings that take place for the $l=1$ p modes.

The frequency spectrum of $l=0$ and $l=1$ modes at any particular stellar age
can be read off 
Fig.~\ref{fig:gemmaevol} by taking a vertical cut through the ridges: an 
example at an age of about 5.98\,Gyr is indicated. It is evident that the 
$l=0$ modes will be essentially evenly spaced in frequency, consistent 
with the asymptotic expression (Eq.~\ref{eq:pasymp}), whereas the 
series of avoided crossings will cause the $l=1$ modes to be nonuniform. 
Moreover, the location in frequency space where avoided crossings are 
``caught in the act'' is strongly dependent on the age of the star and so 
can enable the age of the star to be determined rather precisely.

This behaviour is consistent with measured frequencies of Gemma, illustrated
in the \'echelle diagram in
Fig.~\ref{fig:gemmaechl}. The $l=0$ and $l=2$ frequencies are approximately 
uniformly spaced, whereas the $l=1$ frequencies are more irregularly
spaced, consistent with avoided crossings.
(The $l=2$ and higher-degree p modes are affected much less 
by coupling to the g modes than are the $l=1$ modes, 
because their lower turning points are further from the core and 
hence the evanescent region between the resonant cavities of the p and g modes
is wider.) 

\citet{Metcal2010} modelled Gemma, fitting to the individual measured
frequencies and hence exploiting in particular the non-uniform distribution
of the $l=1$ modes. The frequencies of one of their resulting models are
shown in the \'echelle diagram. They found that two families of solutions, 
one with stellar masses around $1.1\,M_\odot$ and the other with stellar
masses around $1.2\,M_\odot$, %  \note[maybe should say 1.1 and 1.2 instead?], 
that fitted the observed frequencies equally well. Notwithstanding this
10\% ambiguity in mass, the radius and age of the star were determined
with a precision of about 1\%, and an estimated accuracy of about 2\% for the
radius and about 15\% for the age. The mass ambiguity would be resolved if
the range of measured frequencies could be extended to higher frequencies, 
at which point the model frequencies not only of the $l=1$ modes but also
of the $l=0,2$ modes diverge between the two families of models.

\begin{figure}[b]
% \vspace*{-2.0 cm}
\begin{center}
\includegraphics[width=8cm]{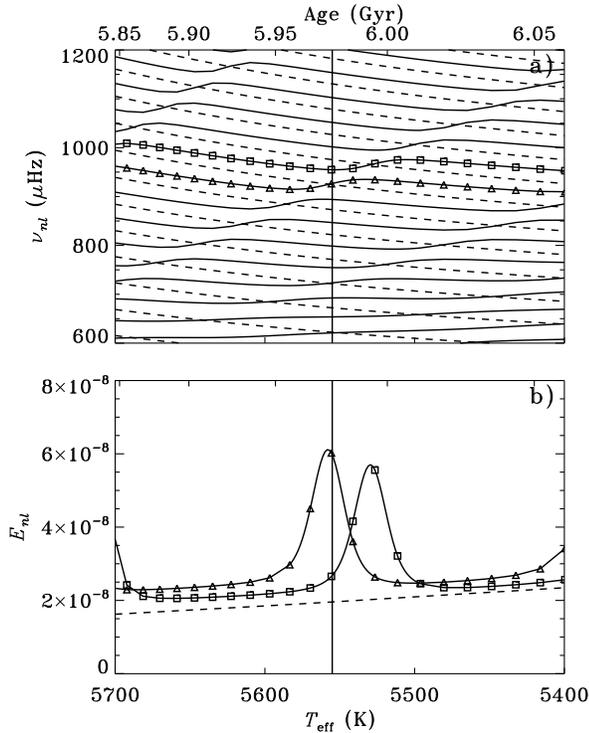} 
%\vspace*{-1.0 cm}
\caption{ Panel a) shows a blow-up of Fig.~\ref{fig:gemmaevol} around
the model illustrated in Fig.~\ref{fig:gemmaechl}.
Panel b) shows mode inertias for two of the $l = 1$ modes
(solid lines, identified in both panels by triangles and squares, respectively),
and the intervening radial mode (dashed line).
The increase in the inertia, relative to the radial mode,
is an indication of a predominant g-mode character of the modes.
%\note [Detailed Gemma frequency and inertia evolution.]
%\note[Not sure we need this figure - we could reference it, though I have not.
%!mjt]
   \label{fig:gemmainertia}}
\end{center}
\end{figure}

The properties of the modes in the vicinity of the avoided crossings is
illustrated in more detail in terms of the mode inertia
(cf.\ Eq.~\ref{eq:inertia}) in Fig.~\ref{fig:gemmainertia}.
When the dipolar modes behave predominantly as acoustic modes,
with frequencies decreasing with increasing age, their inertia is
close to that of the neighbouring radial mode.
It increases when a mode behaves predominantly as a g mode;
at the point of closest approach in an avoided crossing the two modes
have the same inertia, intermediate between the g-mode and the p-mode
behaviour.
As discussed below for red giants, the inertia has an important influence
on the mode visibility.
For the dipolar modes in the Gemma models the contrast between
the p- and g-mode behaviour is modest and the modes are readily observable,
even when they are most g-mode-like.
On the other hand, for modes of degree $l = 2$ and higher the intermediate
evanescent region is substantially broader and the distinction between the
p- and g-mode behaviour correspondingly stronger.
Thus it is less likely to observed mixed modes at these higher degrees,
as confirmed by the interpretation of the observed spectrum.

Apart from the intrinsic interest in the asteroseismic studies of solar-like
stars, these studies provide an important possibility for characterizing 
stars that host extra-solar planetary systems.
When a planet is detected using the transit technique the variation
in the detected stellar brightness depends on the ratio between the diameters
of the planet and the central star.
A reliable determination of the planet radius, of great importance to
the characterization of the nature of the planet, therefore depends on
determining the stellar radius.
As discussed above this can be provided by the asteroseismic analysis
of the oscillations of the star;
indeed, the target stars for the {\it Kepler} search for extra-solar
planets are generally in a range of parameters where solar-like oscillations
are expected, although most stars are too faint for reliable observations
to be possible.
However, the potential of the technique was demonstrated by
\citet{Christ2010} who analysed {\it Kepler} asteroseismic data for
a known exoplanet host.
More recently, asteroseismic analysis was used in the characterization
of the first rocky planet detected by {\it Kepler} \citep{Batalh2011}.

%\note [Make a few comments on the use of asteroseismology to characterize
%planet hosts, with {\it Kepler}, to preface PLATO presentation?]

We finally note that the frequencies of solar-like oscillation are sensitive
to stellar magnetic activity.
In the solar case this has been studied extensively
\citep[e.g.,][]{Woodar1985, Libbre1990}.
Some evidence for such variation was found by \citet{Metcal2007} for the
star $\beta$ Hyi.
As discussed by \citet{Karoff2009} the long observing sequences possible
with CoRoT and in particular
{\it Kepler} provide a rich possibility for detecting similar effects
in other stars.
In fact, in a solar-like star observed by CoRoT
\citet{Garcia2010} detected variations in oscillation frequencies
and amplitudes which appeared to be the result of a rather short
stellar activity cycle.

\section{Red giants}

\label{sec:redgiant}
%\note [Briefly on prediction of stochastic excitation, possibly also results
%for semi-regular variables, including AAVSO result, but now thousands of
%observations, from CoRoT and Kepler.]
%
Assuming that the solar oscillations are excited stochastically by convection
\citep{Goldre1977} one would expect that all stars with vigorous outer 
convection zones exhibit such oscillations.
A rough estimate of the amplitudes \citep{Christ1983} suggested that the 
amplitude increases with increasing luminosity.
Thus red giants are obvious targets for the search for, and investigation of,
solar-like oscillations.

%\note [Early indication in Arcturus.]
%\note [Cluster studies; but they may have been later?]
The first definite detection of individual modes in a red 
giant was obtained by \citet{Frands2002} in the star $\xi$ Hya.
The frequency spectrum showed very clearly a series of uniformly spaced
peaks with a separation $\Delta \nu \simeq 7 \muHz$.
Simple modelling, given also the location in the HR diagram, strongly
suggested that only radial modes had been observed;
the alternative identification, of alternating $l = 0$ and $l = 1$ modes,
would correspond to a true large separation twice as big and hence
a radius smaller by roughly a factor 1.3, entirely inconsistent with the 
observed luminosity and effective temperature.

Evidence for solar-like oscillations in giants had also been obtained from
more statistical analyses.
\citet{Christ2001} noted that the relation between the standard deviation
and mean of the amplitude variations in the so-called semi-regular variables,
based on visual observations carried out by the American Association of
Variable Star Observers (AAVSO), was consistent with the expectations for
stochastically excited oscillations.
The solar-like nature of the oscillations of selected semi-regular 
variables was confirmed by
\citet{Beddin2003b} through analysis of their power spectra.
Also, \citet{Kiss2003, Kiss2004} analysed large sets of OGLE%
\footnote{Optical Gravitational Lensing Experiment}
observations of red giants, 
obtaining clear indication of several modes of oscillation which may
likely be identified as solar-like.
Detailed analyses of the OGLE data were carried out by
\citet{Soszyn2007} and \citet{Dziemb2010}, confirming the solar-like
nature of the observed oscillations.
These investigations extend the realm of solar-like oscillations to stars with
a luminosity of up to $10\,000\, L_\odot$ and periods of several months.

%\note [Possibly a little on clusters here?]

%\note [In the discussion of the oscillation properties, assume that effects 
%of compact core, with mixed modes etc., have been discussed in some detail
%in Section~\ref{sec:solar-like}.]

\begin{figure}[b]
% \vspace*{-2.0 cm}
\begin{center}
\includegraphics[width=4in]{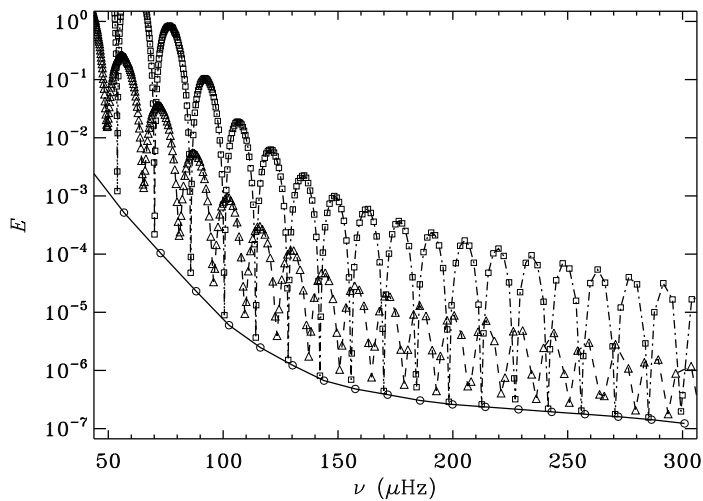} 
%\vspace*{-1.0 cm}
 \caption{Mode inertias (cf.\ Eq.~\ref{eq:inertia}) against cyclic
frequency in a red-giant model of mass $1.4 \, M_\odot$ and
radius $5 R_\odot$.
Modes of degree $l = 0$ (circles connected by a solid line),
$l = 1$ (triangles connected by a dashed line) and
$l = 2$ (squares connected by a dot-dashed line)
are illustrated.}
   \label{fig:inertia}
\end{center}
\end{figure}

\begin{figure}[b]
% \vspace*{-2.0 cm}
\begin{center}
\includegraphics[width=4in]{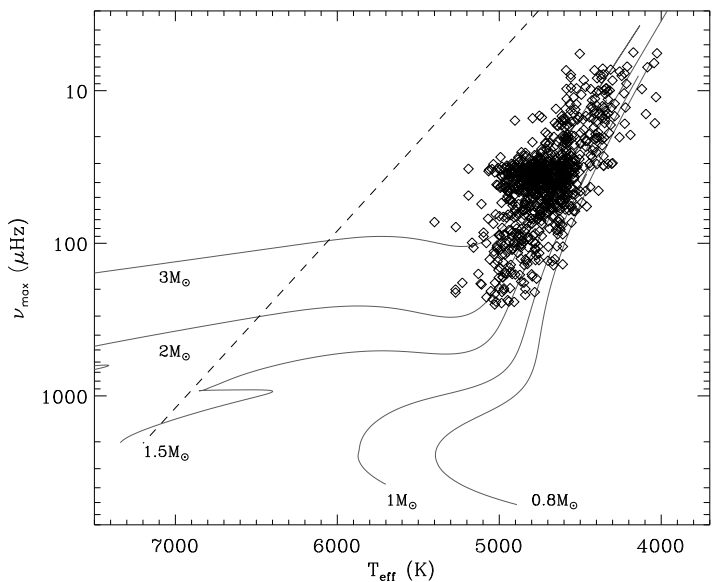} 
%\vspace*{-1.0 cm}
 \caption{`Hertzsprung-Russell diagram' of red giants observed with
{\it Kepler}, using the frequency $\nu_{\rm max}$ at maximum power
as a proxy for luminosity.
The curves show evolution tracks for models at the indicated ages.
Figure courtesy of D. Huber \citep[see][]{Huber2010}.}
   \label{fig:huber_hr}
\end{center}
\end{figure}

%\note [Here or earlier need reference to a general review on red giants.]
The red-giant phase \citep[see][for a review]{Salari2002}
follows after the phase exemplified by Gemma, discussed
in Section~\ref{sec:solar-like} above.
The stars ascend the Hayashi region at almost constant effective temperature
and strongly increasing radius and luminosity,
with a very compact helium core and an extended, mostly convective,
envelope.
The energy production takes place through hydrogen fusion in a thin shell 
around the helium core.
The tip of the red-giant branch is defined by the ignition of helium near the
centre.
%\note [May or may not mention possibility of He flash.]
Stars with central helium fusion are located in the so-called `red clump' 
in the HR diagram (see Fig.~\ref{fig:huber_hr});
even for these, however, most of the energy is produced by the hydrogen shell.
The strongly centralized helium fusion gives rise to a small convective core,
although the bulk of the helium core remains radiative.
In both the ascending red giant and the clump phase the small extent of the
core gives rise to a very high gravitational acceleration and hence 
buoyance frequency, further amplified by the presence of strong 
composition gradients (cf.\ Eq. \ref{eq:buoy}).
Thus all nonradial modes have the character of high-order g modes in the core.
%\note [The discussion here of mode inertia
%probably follows naturally from discussion in Section~\ref{sec:solar-like}.]
The resulting mixed nature of the modes, and the high density of modes
of predominantly g-mode character, are illustrated in Fig.~\ref{fig:inertia}
which shows the mode inertia $E$ (cf.\ Eq.~\ref{eq:inertia}) for a
typical model on the ascending red-giant branch.
Most of the modes with $l = 1$ and $2$ clearly have much higher inertias than
the radial modes, and hence are predominantly g modes.
However, there are resonances where the modes are largely trapped in
the outer acoustic cavity. 
These p-dominated modes have inertias close to the inertia of the radial
modes, reflecting their small amplitudes in the core.

\citet{Dziemb2001} considered the excitation and damping of modes
in red giants.
They found that the very high radial order of the nonradial modes in the core
led to substantial radiative damping, even for modes that were predominantly
acoustic.
%\note [need to check precisely how Dziembowski et al. phrase this.]
On this basis \citet{Christ2004} concluded that nonradial oscillations 
were unlikely to be observed in red giants.
(This would be consistent with the results of \citet{Frands2002} on
$\xi$ Hya where apparently only radial modes were found.)
Fortunately this conclusion was wrong: nonradial modes are indeed observed
in red giants and provide fascinating diagnostics of their interiors.

\begin{figure}[b]
% \vspace*{-2.0 cm}
\begin{center}
\includegraphics[width=2in]{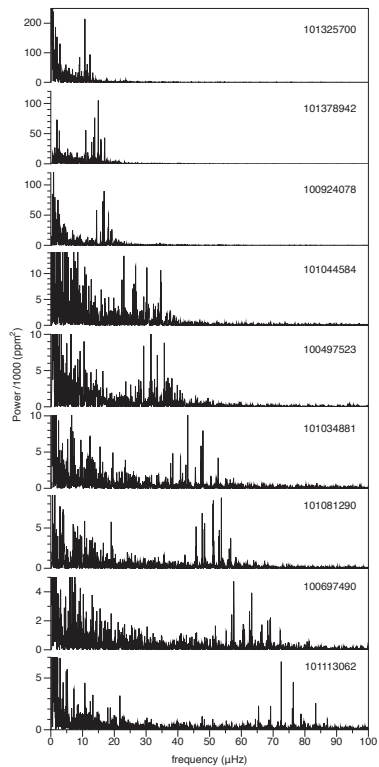} 
%\vspace*{-1.0 cm}
 \caption{Power spectra of solar-like oscillations in red giants,
from five months of observations with CoRoT.
The stars are identified by their CoRoT identification number,
with radius increasing towards the top.
{}From \citet{DeRidd2009}.}
   \label{fig:corotrg}
\end{center}
\end{figure}

A first indication of the presence of nonradial oscillations in red giants
came from line-profile observations by \citet{Hekker2006}.
%\note [need to check precisely what was concluded.
%Also discuss MOST observations of $\epsilon$ Oph, nonradial modes identified
%by Kallinger??]
However, the major breakthrough came with the observation of a substantial
number of red giants by CoRoT, as presented by \citet{DeRidd2009}.
A selection of the resulting power spectra are shown in Fig.~\ref{fig:corotrg}.
The presence of solar-like oscillations is obvious, the peaks shifting to lower
frequency with increasing radius (cf.\ Eq.~\ref{eq:numax}).
Also, \citet{DeRidd2009} showed an example of an \'echelle diagram
which beyond any doubt identified modes of degree 0, 1 and 2.

%\note [Would be nice to show plots from De Ridder et al., Nature, 2009, if
%possible, as the main breakthrough.]

The potential visibility of nonradial modes in red giants
was made very much clearer by \citet{Dupret2009},
following an analysis by \citet{Chapli2005}.
The observational visibility of a mode is determined by the peak
height $H$ in the power spectrum.
This is related to the total observed power $P$ of the mode
by $P \propto H \Delta$, where $\Delta$ is the width of the peak.
If the mode is observed for much longer than the natural damping time,
the width is given by the damping rate, i.e., the imaginary part
$|\omega_{\rm i}|$ of the frequency.
If the damping is dominated by the near-surface layers, as is often the
case, at a given frequency $\omega_{\rm i}$ is related to the mode inertia $E$
by
\begin{equation}
\omega_{\rm i} \propto E^{-1} \; .
\label{eq:omegai}
\end{equation}
Thus those modes that are predominantly g modes, with high inertia
(cf. Fig.~\ref{fig:inertia}) have much smaller widths than 
the p-dominated modes.
The power in the mode is determined by a balance between the energy 
input from the stochastic noise near the stellar surface and the damping.
Assuming again Eq.~(\ref{eq:omegai}) the outcome is that $P \propto E^{-1}$
at fixed frequency.
It follows that the peak height $H$ is independent of $E$ at a given frequency
and hence that the g-dominated modes should be observed at the same height
as the p-dominated modes.

This, however, assumes that the duration ${\cal T}$ of the observation is
much longer than the lifetime $|\omega_{\rm i}|^{-1}$ of the mode.
If this is not the case, the peaks are broader and the height consequently
smaller.
As an approximate scaling of this dependence \citet{Fletch2006}
proposed
\begin{equation}
H \propto {P \over |\omega_{\rm i}| + 2/{\cal T}} \; ;
\label{eq:height}
\end{equation}
for ${\cal T} \ll |\omega_{\rm i}|^{-1}$,
in particular, 
$H \propto P \propto E^{-1}$
and the g-dominated modes are essentially invisible.
As is clear from Fig.~\ref{fig:inertia} the g-dominated modes in practice often
have inertias much higher than the p-dominated modes and hence
correspondingly longer lifetimes;
thus they may be expected
have small observed peak heights, unless observations of very
long duration are analysed.
However, modes of mixed character, particularly those with $l = 1$,
may have damping times comparable with or shorter than the very long
observations made available by CoRoT and {\it Kepler} and hence may be visible.
Concerning $\xi$ Hya,
the apparent absence of nonradial modes in the observations
was probably caused by the 
relatively short observing run of around one month,
compared with the five-month observations by \citet{DeRidd2009}.
Even the most p-mode-like dipolar modes have somewhat higher mode inertia 
and hence lifetimes than the radial modes; thus the peak height of these modes 
was likely suppressed in the observations by \citet{Frands2002}.

\begin{figure}[b]
% \vspace*{-2.0 cm}
\begin{center}
\includegraphics[width=4in]{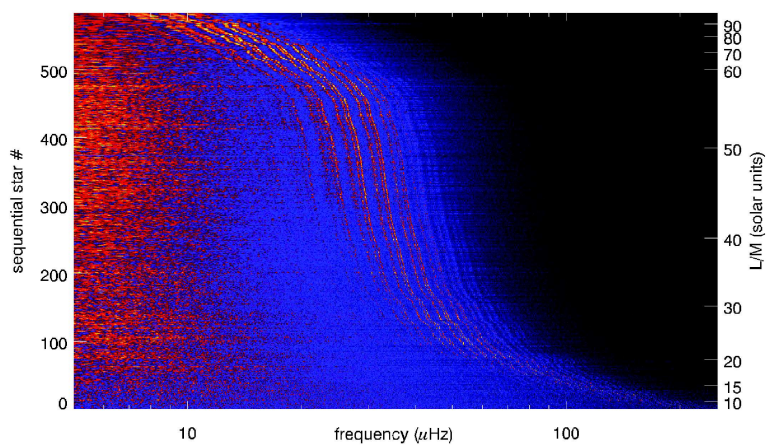} 
%\vspace*{-1.0 cm}
 \caption{Stacked power spectra of red giants observed with {\it Kepler},
with the frequency of maximum power decreasing towards the top.
The numbers at the right edge provide an estimate of the corresponding
ratio between luminosity and mass, in solar units.
Figure courtesy of T. Kallinger.
}
   \label{fig:stackspec}
\end{center}
\end{figure}

%\note [Probably show stacked power spectra as in Gilliland et al., updated?
%Against inferred radius or luminosity?]

Very extensive results on red-giant oscillations have been obtained by
CoRoT and {\it Kepler}
\citep[e.g.,][]{Hekker2009, Beddin2010, Mosser2010, Kallin2010a, Kallin2010b,
Stello2010, Hekker2011}. 
These confirm the acoustic nature of the observed spectra, 
with a clear detection of modes of degree 0, 1 and 2.
This is illustrated in Fig.~\ref{fig:stackspec} 
\citep[see also][]{Gillil2010},
for a large sample of stars 
observed with {\it Kepler} during the first 16 months of the mission;
the observations are shown as stacked spectra ordered according
to decreasing large 
frequency separation $\Delta \nu$ and hence increasing radius and luminosity.
This is characterized at the right-hand edge of the figure
by the ratio between luminosity and mass,
estimated from the oscillation parameters (see below).
Stellar `noise' from granulation is evident in the low-frequency region.
%This clearly shows a tight relation between the frequency $\nu_{\rm max}$ at
%maximum power and the luminosity \note [which may have been discussed already].
The frequencies approximately satisfy an asymptotic relation similar to 
Eq.~(\ref{eq:pasymp}),
with a closely spaced pair of bands of $l = 0, 2$ modes and an intermediate
band of $l = 1$ modes.
However, since the acoustic propagation region is generally confined to
the convection zone or the region just beneath it
the small separation between $l = 0$ and 2 is not directly related to the 
properties of the stellar core, let alone the age of the star, unlike
the situation on the main sequence.
\citet{Montal2010a} carried out an extensive analysis of the overall properties
of the oscillation frequencies for a large sample of red-giant models.
They noted that the outer convective zone in red-clump phase is not quite as
deep as for the stars ascending the red-giant branch;
the extent of the acoustic propagation region beyond the convective
envelope was found to have a
potentially measurable effect on the small frequency separations.

The mean large frequency separation $\Delta \nu$ and the frequency 
$\nu_{\rm max}$ at maximum power satisfy the scaling relations
(\ref{eq:nuscale}) and (\ref{eq:numax}).
Thus the stellar properties can be characterized by these quantities.
This is used in Fig.~\ref{fig:huber_hr} to plot a `Hertzsprung-Russell'
diagram of red giants observed with {\it Kepler}, replacing the
luminosity by $\nu_{\rm max}$ as a measure of radius and hence luminosity.
The observations are compared with evolution tracks for a range of masses,
using scaling from the solar value of $\nu_{\rm max}$.
The distribution of stars clearly shows the higher density in the region 
of the helium-burning red clump.
The scaling relations can also be used to determine the stellar parameters
from the observed $\Delta \nu$, $\nu_{\rm max}$ and $T_{\rm eff}$
\citep[e.g.][]{Kallin2010a}.
This provides a unique possibility for population studies of red giants
\citep[e.g.,][]{Miglio2009, Kallin2010b, Mosser2010, Hekker2011}.
The CoRoT results are particularly interesting in this regard,
given that they allow a comparison of the populations in the centre and
anti-centre directions of the Galaxy (Miglio et al., in preparation)
and hence provide information about the evolution and dynamics of the
Galaxy.
A more precise determination of the stellar parameters can be obtained
with the so-called grid-based methods, where stellar modelling is used
to relate the effective temperature, mass and radius
\citep[e.g.,][]{Gai2011}.
This was used by \citet{Basu2011} to investigate the properties of
two open clusters in the {\it Kepler} field.

\begin{figure}[b]
% \vspace*{-2.0 cm}
\begin{center}
\includegraphics[width=2.5in]{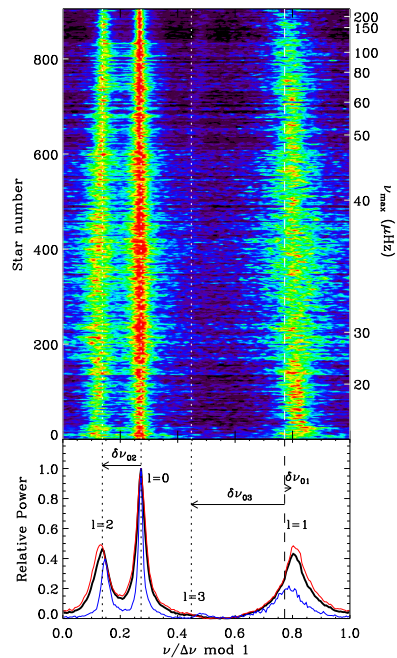} 
%\vspace*{-1.0 cm}
 \caption{The upper panel shows stacked collapsed, rescaled and
shifted \'echelle diagrams for red giants observed with {\it Kepler}.
These have been collapsed in the lower panels, where the 
peaks corresponding to $l = 0 - 3$ are indicated.
The thick lines correspond to the full set, and the thin blue and red lines
(in the electronic version)
correspond to stars with $\nu_{\rm max} > 100 \muHz$ and 
$\nu_{\rm max} < 50 \muHz$, respectively.
Figure courtesy of D. Huber \citep[see][]{Huber2010}.}
   \label{fig:huber_echl}
\end{center}
\end{figure}

To the extent that the modes are trapped in the convective envelope,
whose structure is very similar amongst different stars
apart from a scaling depending on the stellar mass and radius, one would
expect a corresponding similarity between the oscillation frequencies.
This is confirmed by the so-called `universal red-giant oscillation pattern',
a term introduced by Mosser et al., of the oscillations
\citep{Huber2010, Mosser2011}.
To illustrate this, Fig.~\ref{fig:huber_echl} shows a
normalized and stacked \'echelle diagram.
%\note [of the stars shown in the previous figure?].
Here collapsed \'echelle diagrams for the individual stars,
normalized by the large separation,
have been stacked after taking out the variation with stellar parameters
of the average $\epsilon$ (cf.\ Eq. \ref{eq:pasymp}).
Clearly there is very little variation with $\nu_{\rm max}$
and hence stellar properties
in the location of the ridges and hence the scaled small separations.
This is emphasized by the collapsed version of the diagram in the lower panel;
it should be noticed that this also, as indicated, provides a weak
indication of modes of degree $l = 3$.
A lower limit to the width of the ridges is provided by the 
natural width of the peaks, corresponding to the lifetime of the modes.
For $l = 0$ and 2 \citet{Huber2010} found a width of around 
$0.2 \muHz$, essentially independent of the stellar parameters and
corresponding to a mode lifetime%
\footnote{defined as the e-folding time of the displacement}
of around 18\,d.
Similar results were obtained by \citet{Baudin2011}
based on CoRoT observations.
Interestingly, this value of the lifetime is similar to the 
estimate obtained by \citet{Houdek2002} from modelling of $\xi$ Hya.
%\note [Say something about width and hence damping time; return to 
%broader dipolar modes below.]

In Fig.~\ref{fig:huber_echl} it is evident that the ridge for
$l = 1$ appears substantially wider than for $l = 0$ and 2.
This can be understood from the analysis of \citet{Dupret2009},
discussed above, which showed that several dipolar modes may reach
observable amplitudes in the vicinity of an acoustic resonance
\citep[see also][]{Montal2010b}.
%\note [Needs a little coordination with the above discussion.]
In a major breakthrough in red-giant asteroseismology
\citet{Beck2011} and \citet{Beddin2011} demonstrated 
that the frequencies in such groups of peaks showed clear 
g-mode-like behaviour; this allowed a determination
of the uniform g-mode period spacing (cf.\ Eq.~\ref{eq:gper}).
It was further demonstrated by \citet{Beddin2011} that the 
inferred value of $\Delta \Pi$ allowed to distinguish between stars on
the ascending red-giant branch and stars in the helium-burning 
clump phase.
With further analyses these observations will undoubtedly provide 
very valuable diagnostics of the central properties of red giants.

%\note [Possibly HR diagram in terms of $\nu_{\max}$.]

%\note [Ideally histogram of mass distribution in centre and anti-centre
%directions.]

The frequencies of purely acoustic modes also contain information beyond the 
basic parameters $\Delta \nu$ and $\nu_{\rm max}$.
Sharp features in the sound speed lead to systematic departures from
the simple asymptotic behaviour in Eq.~(\ref{eq:pasymp}), with 
characteristic properties which provide information about the location and
strength of the feature \citep[e.g.,][]{Gough1990}.
An important example is the effect on the sound speed of the localized
reduction in the adiabatic exponent caused by the second ionization of
helium, which provides information about the helium abundance
\citep[e.g.,][]{Voront1991, Montei2005, Houdek2007}.
\citet{Carrie2010} and \citet{Miglio2010} found the signature of this
effect in the red giant HR~7349 observed by CoRoT.
Miglio et al.\ noted that the inferred location of the second helium
ionization zone provided a strong constraint on the properties of the star.
Such analyses will undoubtedly be possible for a large number of red
giants observed by CoRoT and {\it Kepler}.

\section{Concluding remarks}

%\note [CoRoT continuing, just at beginning of {\it Kepler} analysis.
%SONG is coming, and we hope for PLATO.
%\note [Link to Juri's growing interest in stellar modelling,
%including convective
%cores and stellar dynamos, and hope to draw on his organizational skills in
%these new projects.]
%
The last few years have seen amazing progress in the availability of data
for asteroseismic investigations of stellar interiors.
The full scientific exploitation of these data is just starting.
The community is realizing the challenges provided by actual observations,
compared with the simulations that preceded the missions,
and we are hard at work at
optimizing the techniques for the data analysis and interpretation, with the
goal of addressing specific aspects of stellar interiors.
The ability to study solar-like oscillations in hundreds of stars with
{\it Kepler} has been
a positive surprise, allowing comparative studies in
{\it ensemble asteroseismology}%
\footnote{or, according to D.~O.\ Gough, perhaps better
{\it synasteroseismology}}
\citep{Chapli2011};
however, we still need to extend the investigations
to unevolved stars of lower masses, as will surely be possible as ever
longer timeseries for individual stars are accumulated.
The present authors, at least, had not foreseen the possibility 
of detailed analyses
of the mixed modes in subgiants, providing archaeological information about 
the properties of the mixed cores during the main-sequence phase
\citep{Deheuv2010}.
The very recent detection and analysis of features in the pulsations
directly related to the g modes in the cores of red giants
\citep{Beck2011, Beddin2011} have also been
a major surprise, with huge potentials both for characterizing the evolutionary
state of the stars and for investigating the properties of the deep 
interiors of these stars.
The detection in the {\it Kepler} field of a dozen subdwarf B stars 
showing long-period oscillations, with additional cases being found 
by CoRoT, is providing tight constraints on the overall properties of stars
in this very late phase of evolution and there is certainly a potential for
much more detailed investigations.
And the list goes on.

We are fortunate currently to have access to three largely complementary
space missions with asteroseismic potentials.
The MOST mission is perhaps somewhat overshadowed by CoRoT and {\it Kepler},
but it continues to provide excellent data on a variety of bright stars
with the possibility of selecting targets over a large part of the sky;
the recent combined analysis of MOST data and ground-based radial-velocity
data for Procyon \citep{Huber2011} demonstrates the potential of MOST
even in the area of solar-like pulsations.
CoRoT has fully demonstrated the required sensitivity to study
solar-like oscillations in main-sequence stars.
The mission has the very substantial advantage of being able to observe both in
the direction towards and away from the Galactic centre which allows 
comparative studies of stellar populations.
Also, the stars observed in the asteroseismology field are relatively bright,
facilitating the ground-based support observations of these targets,
and the CoRoT `eyes' contain a very broad sample of interesting targets
of most types.
Finally, {\it Kepler} can observe stars for the duration of the mission, 
optimizing the precision and sensitivity of the observations and allowing
the uninterrupted study of potential changes in stellar properties.
The heliocentric orbit provides a more stable and quiet environment 
than the low-earth orbit of CoRoT, in terms of scattered light and 
magnetospheric disturbances.
Also, the {\it Kepler} field has been found to be extremely rich in a variety
of interesting stars, now to a large extent characterized through the 
KAI survey phase.

Even so, there is a continued need to improve the observational situation
and strong prospects that this will happen.
The BRITE Constellation mission,
under development by Austria, Canada and Poland \citep{Kuschn2009}
will fill a very important niche by carrying out photometric observations
of the brightest stars across the sky in two colours.
On a longer timescale the ESA PLATO mission, if selected, will
greatly extend the {\it Kepler} results \citep{Catala2011}.
As {\it Kepler}, PLATO has the dual purpose of exoplanet investigations
and asteroseismology.
However, PLATO will look at substantially brighter stars in much larger fields.
This is important for the crucial ground-based follow-up observations to
confirm the detection of a planet in an apparent transit, particularly 
for earth-size planets which will be a key goal for PLATO as it is for
{\it Kepler}.
Also, as a result PLATO will allow asteroseismic characterization of
a substantial fraction of the stars around which potential planets are
found, unlike {\it Kepler} where this is an exception.
PLATO will be placed in an orbit around the ${\rm L_2}$ point which 
shares the advantage, in terms of stability, of {\it Kepler's} 
heliocentric orbit.
The planned observing schedule consists of continuous observations of two 
fields for two or three years each, followed by a `stop-and-stare' phase
where each field is observed for a few months.
The latter part of the mission will allow investigation of a substantial
fraction of the sky, providing a survey of a far more wide-ranging and varied
set of stellar pulsations and other types of stellar variability than has
been possible even with {\it Kepler}. 

The great advances provided by the space asteroseismic observations
should not blind us to the continued usefulness of ground-based
observations.
In photometry that allows study of rare objects that may not be available
to the space observatories.
Also, it provides greater flexibility, e.g., in carrying out observations
in several colours, of importance to mode identification.
Even more important is the use of ground-based radial-velocity observations,
particularly for solar-like oscillations.
Observations of solar oscillations from the SOHO mission have clearly
demonstrated that the solar background, from granulation and activity,
is substantially higher relative to the oscillations for photometric 
observations than for radial-velocity observations,
as already noted by \citet{Harvey1988}
\citep[see also][]{Grunda2007}.
This background is evident in the {\it Kepler} observations of Gemma
illustrated in Fig.~\ref{fig:gemmaspec}.
This puts a natural limit to the precision and mode selection possible
in photometric observations.
Thus radial-velocity observations of carefully selected stars are still
required to reach the ultimate precision and level of detail in asteroseismic
investigations.
Such observations can been carried out from the ground, as has been
done successfully for a small sample of stars 
\citep[e.g.,][]{Bouchy2002, Beddin2004, Bazot2005, Kjelds2005, Beddin2007,
Arento2008};
to reduce the gaps in the data they have been carried out in a coordinated
fashion, involving two or more observatories.
However, the required state-of-the-art instrumentation is only available
for limited periods and certainly not for the month-long observations
from several observatories that
are needed to secure adequate frequency resolution.
This is the motivation for the development of the Stellar Observations
Network Group (SONG) network \citep{Grunda2009, Grunda2011}.
%\note [Possibly include SONG logo as a figure; probably not more.]
SONG is planned to consist of 7 -- 8 nodes with a suitable geographical
distribution in the northern and southern hemisphere.
Each node will consist of a 1\,m telescope, equipped with a high-resolution 
spectrograph for Doppler-velocity observations and a so-called lucky-imaging
camera for photometry in crowded fields.
With the use of an iodine cell as reference, and with careful optimization
of the optics, it is estimated that SONG will be able to study 
velocity oscillations in a star as the Sun at magnitude 6.
The lucky-imaging camera is designed for characterization of exoplanet 
systems through gravitational micro-lensing \citep[e.g.,][]{Domini2010}.
At present a prototype SONG node is under construction with Danish funding,
to be placed at the Iza{\~n}a Observatory on Tenerife,
with expected deployment and start of operations in 2011.
A Chinese node is being designed and is expected to be operational in 2013,
and funding and support for further nodes will be sought through a network
of international collaborations.

The data from these projects will provide excellent possibilities for testing
present stellar evolution calculations and indicating where improvements 
should be made.
Such improvements are certainly required, particularly when it comes to the
treatment of stellar internal dynamics.
Impressive progress is being made in the application of complex, yet 
unavoidably highly simplified, treatments of the interplay between rotation,
circulation, and magnetic fields, 
including also the evolution of stellar internal angular velocity 
\citep[e.g.,][]{Palaci2003, Palaci2006, Mathis2004, Mathis2005}
\citep[see also][]{Maeder2009}.
Indeed, full stellar evolution calculations will undoubtedly require such
simplifications in the foreseeable future.
However, it is equally important that these simplified treatments be tested
by detailed simulations of the hydrodynamical phenomena, albeit for conditions
that often do not fully reflect the stellar internal conditions.
An important example is the simulation of near-surface convection, where
computations under reasonably realistic conditions are in fact possible
\citep[e.g.,][]{Nordlu2009, Trampe2010}.
%\note [Nordlund et al. reviews; perhaps a suitable Trampedach paper,
%with a few words].
Simulations of the deeper parts of the convective envelope and the region
below \citep[see][for reviews]{Miesch2005, Miesch2009}
unavoidably require simplification, but are providing deep insights
into the interaction between convection and rotation, and the generation
of magnetic fields \citep{Brun2004, Browni2006, Miesch2008}.
%\note [several Toomre et al. references, perhaps Miesch LRSP review].
%It is very encouraging that 
Such simulations are now being extended to
stars other than the Sun \citep{Brun2009, BrownB2011},
including the dynamics of stellar convective cores 
\citep{Brun2005, Feathe2009}.
The observations by CoRoT, {\it Kepler} and projects to follow provide excellent
prospects for testing the results of such modelling and hence improve our
understanding of stellar internal dynamics.

\medskip\noindent
{\it Acknowledgement}: We wish to take this occasion to thank Juri for many 
years of enjoyable collaboration, as well as for his inspiration and 
constant friendship.
We are very grateful to 
P. Degroote,
J. De Ridder,
G. Do{\u g}an,
D. Huber,
T. Kallinger,
C. Karoff, 
K. Kolenberg,
R. Szab\'o and
V. Van Grootel
for the provision of, or help with, the figures, and to Travis Metcalfe for
comments that helped improve the paper.
We thank Sacha Brun and Nick Brummell for the excellent organization of an
exciting conference, and for their patience with the present authors.
The National Center for Atmospheric Research is sponsored by the 
National Science Foundation (NSF).

%\begin{discussion}
%
%\discuss{J. Toomre}{ 
%Could you explain of what have been the surprises of pulsation results
%for red-giant stars?
%}
%
%\discuss{Name}{ 
%... answer ...
%}
%
%\discuss{E. Zweibel}{ 
%If you interpret the $T_{eff}$ in the $A$--$L/M$--$T_{eff}$ relation
%as the Hayashi temperature can you interpret the relation theoretically?
%}
%
%\discuss{Name}{ 
%Insofar as I understood the answer, it has to do with the scaling of
%convection amplitude with respect to $L$, as a relation between mode
%power and convection power.
%}
%
%%%%%%%%%%%%%%%%%%%%%%%%%%%%%%%%%%%%%%%%%%%%%%%
%\end{discussion}

\end{document}